\documentclass[twocolumn,showpacs,preprintnumbers,prl,amsmath,amssymb]{revtex4-1}
\usepackage[colorlinks=true,citecolor=blue,linkcolor=magenta]{hyperref}
\usepackage{graphicx}
\usepackage{dcolumn}
\usepackage{bm}
\usepackage{bbm}
\usepackage[english]{babel}
\usepackage{color}
 
\begin{document}
\title{Nanodiamond Collective Electron States and their Localization}
\author{Ivan A. Denisov$^{1}$, Andrey A. Zimin$^{1}$, Leslie A. Bursill$^{2}$, Peter I. Belobrov$^{1,3}$}
\email{d.ivan.krsk@gmail.com}
\affiliation{$^{1}$MOLPIT, Institute of Fundamental Biology and Biotechnology,\\Siberian Federal University, 660041 Krasnoyarsk, Russia \\
$^{2}$School of Physics, The University of Melbourne, PO BOX 339, Carlton North VIC 3054, Australia \\
$^{3}$Kirensky Institute of Physics \& Institute of Biophysics SB RAS, 660036 Krasnoyarsk, Russia}

\begin{abstract}
The existence and localization of collective electron states for nanodiamond particles were studied both by solving a one-particle one-dimensional Schr\"odinger equation in the Kronig--Penney potential and by \textit{ab initio} computations of ground state wavefunctions of diamondoids $C_{78}H_{64}$, $C_{123}H_{100}$ and $C_{211}H_{140}$ at the DFT~R-B3LYP/6-31G(d,p) level of theory. Three distinct classes of collective electron states have been found: collective bonding orbitals resembling the morphology of 3D-modulated particle in a box solutions; surface-localized non-bonding conductive Tamm states and subsurface-localized bonding states for non-uniformly compressed nanodiamond. Quantum-mechanical analysis shows that collective unpaired electrons are intrinsic to nanodiamond. Their subsurface localization is described in terms of surface compression arising from a self-consistency condition of the electron-nuclear wavefunction. Intrinsic spin existence is supposed to result from the collective and spread nature of subsurface orbitals, allowing spin-density fluctuation effects to become significant on this length scale. Suggested model allows to explain free spins of nanodiamond exhibited in experiments.
\end{abstract}

\maketitle

\section{Introduction}

Collective electron states are widely used in the theory of metals~\cite{Abrikosov1972}, while bulk dielectric and semiconductor materials are equally well modeled as a rigid network of covalent bonds incorporating lattice defects. Success for classic models is mainly due to extensive use of translational symmetry, which is definitely not the case for nano-sized systems: nanoparticles are mostly imperfections with perfect regions rather than perfect crystals with imperfections. Lattice termination is the main imperfection leading to the consideration of the collective surface-localized electron states problem. However, one should clearly distinguish collective states in a metal and in a dielectric material --- collectivity doesn't necessarily imply conductivity, and there is no contradiction between the ``bonding'' and ``collective'' terms. It's crucial to emphasize unambiguous meaning of the term ``collective'' we use in our paper: collective states in dielectric materials are bonding molecular orbitals, localized over several atomic cores. The fact that collective states despite their importance are rarely used in dielectric materials theory results in seemingly unresolvable issues like inability to describe the nature of the intrinsic spin found in the nanodiamond.

B.~Pate has summarized study of the electronic and atomic structure of the bulk diamond in his classic paper~\cite{Pate1986}. Almost simultaneously the nanodiamond was discovered, and its practical applications attracted great attention~\cite{Yang2002, Butler2008}, leaving behind detailed investigation of the underlying structural and electronic aspects. Nanodiamond-based materials are rapidly growing in number, triggering progress in the understanding of the CVD synthesis methods~\cite{Butler2009} and in the thermophysical applications of nanodiamond-based materials~\cite{Balandin2011}. Despite great success in nanodiamond applications, there is still no in-depth understanding of its structure. Collective electron states are vital part of the global perspective which interrelates consistently magnetism, surface conductivity and the set of nanodiamond experimental ``anomalies''.

The first theoretical investigation of the collective electron states in a bulk diamond was done by I.\,E.~Tamm~\cite{Tamm1932g}. His calculations were based on the idea of electron confinement between periodic lattice potential and vacuum. We have followed Tamm approach assuming that electron could be confined between vacuum and semi-periodic diamond ball kernel.

J.~Kouteck\'y then investigated surface states of a one-dimensional crystal in connection with penetration of surface potential perturbation into the crystal and was the first to introduce the term ``subsurface states''~\cite{Koutecky1957}. P.~Phariseau then studied energy spectrum of amorphous substances and subsurface states in deformed one-dimensional crystals, treating deformed region as amorphous phase~\cite{Phariseau1960}. Direct experimental evidence of the surface layer deformation in nanodiamond molecules follows from PEELS~\cite{Bursill2001, Peng2001}, NMR~\cite{Fang2009} and Auger~\cite{Belobrov2003} data and allows us to make the assumption about subsurface states existence in nanodiamond.

We have concentrated our attempts on study of the collective electron states in nanodiamond in order to relate great amount of well-documented reliable experimental data with the \textit{ab initio} computations of its electronic structure. It was recently demonstrated that it's vital to take nuclear movement explicitly into account in order to describe optical gap of diamondoids~\cite{Patrick2013}. However, full-scale \textit{ab initio} geometry optimization is still impossible for big diamondoids. The optimization procedure must be consistent with the experimentally observed deformation, so we introduce an approximate parametric non-uniform compression to the diamond ball model, which would allow us to skip the geometry optimization and to catch the qualitatively correct picture.

The problem of collective electron states localization in nanodiamond is considered, and qualitative explanation of the nature of the experimentally observed compression, the intrinsic unpaired spin and the PEELS pre-peak is proposed.

\section{Methods of nanodiamond electronic structure investigation}

Orbital localization in nanodiamond was studied on model systems by exact numerical integration of Schr\"odinger equation for one-particle model in one-dimensional Kronig--Penney potential and by \textit{ab initio} computations of ground-state wavefunctions of small diamondoids. Simultaneous usage of both methods allows to establish their ranges of applicability and to give better approximation to real 4--6\,nm nanodiamonds.

\subsection{1D collective electron states analysis}

\begin{figure}[t]
	\begin{center}
	\includegraphics[width=0.44\textwidth]{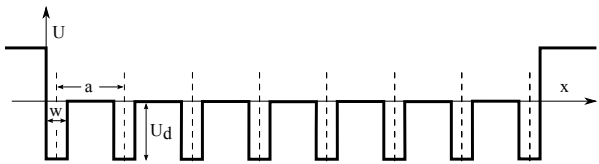}
	\caption{Potential of limited one-dimensional crystal in Kronig--Penney approximation}
	\label{fig:penneyPotential}
	\end{center}
\end{figure}

A one-dimensional one-electron model allows to analyze wavefunctions of big diamond balls (5\,nm and bigger). This model gives clear outcome and its solutions are exact. Despite relative simplicity of one-electron approximations it is widely used in classical solid state physics. Our calculations are made following the logic of I.\,E.~Tamm~\cite{Tamm1932g} and A.\,A.~Abrikosov \cite{Abrikosov1972} using the test electron approach and the model of the Kronig--Penney (Fig.\,\ref{fig:penneyPotential}) for one-dimensional dielectric crystal, bounded on both sides by vacuum. We consider intermediate case between pure dielectric potential in tight-binding approximation and metallic lattice potential were ``free'' electron gas moves in atomic core potential.

Special application was developed to explore the one-dimensional stationary  Schr\"odinger equation solutions in a limited quasi-periodic potential using Component Pascal programming language in the BlackBox Component Builder (Z\"urich, Switzerland)~\cite{Pfister2001}.

\subsection{Design of 3D models of diamond balls}

A set of diamond ball structures in a format compatible with a quantum-chemical software is necessary before one can perform numerical computations of wavefunctions and energies in the three-dimensional case. Application DiaBall (\textit{http://diaball.molpit.com}) was developed to construct diamond balls up to 10\,nm size, visualize and export molecular model in the PDB format. The value of the lattice constant is taken to be $0.35669$\,nm~\cite{Lide2009} in order to build carbon diamond lattice inside defined radius. Two different unit cell arrangements (atom-centered or volume-centered) are possible, varying diamond ball structure slightly.

The structure of the compressed layer is usually and erroneously understood in terms of $sp^2$-shell coating diamond core, despite the vast amount of experimental data showing inadequacy of this model. Thus, NMR spectroscopy is unable to detect a $sp^2$ phase signal~\cite{Fang2009} above the detection 
limit; neither can X-ray powder diffraction. We conclude that the compressed layer is not to do with $sp^2$-phase necessarily, so every atom in our model (and in real best quality nanodiamond, actually) is $sp^3$-hybridized.

Software realization of controlled compression of built structures is characterized by parameters $a$ and $s$ according to the next function:
\begin{equation}
	d r (r) = \dfrac{a}{\left(s - \dfrac{r s}{R}\right)^{\!2} + 1},
	\label{eq:deformation}
\end{equation}
where $d r$ is the atom position shift relative to the origin, $r$ is the distance from atom to the origin, $R$ is the radius of molecule, $a$ is the parameter regulating amplitude of shift and $s$ is the parameter regulating shift decay rate inside diamond ball.

\begin{figure*}[t]
\begin{center}
\parbox[t]{0.33\textwidth}{\center\textbf{a}}\hfill
\parbox[t]{0.30\textwidth}{\center\textbf{b}}\hfill
\parbox[t]{0.30\textwidth}{\center\textbf{c}}
\raisebox{0.25\height}{\includegraphics[width=0.33\textwidth]{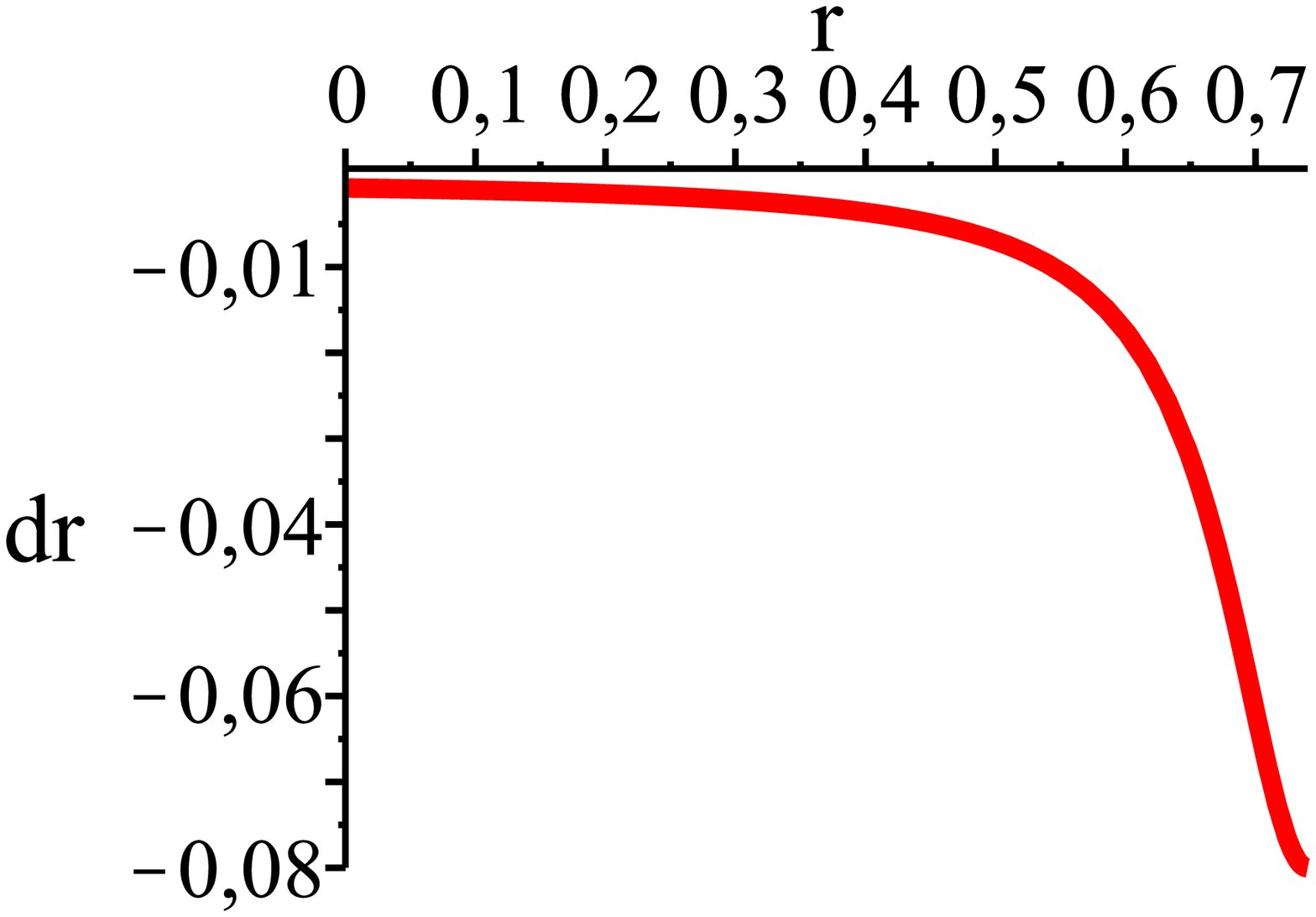}}\hfill
\includegraphics[width=0.30\textwidth]{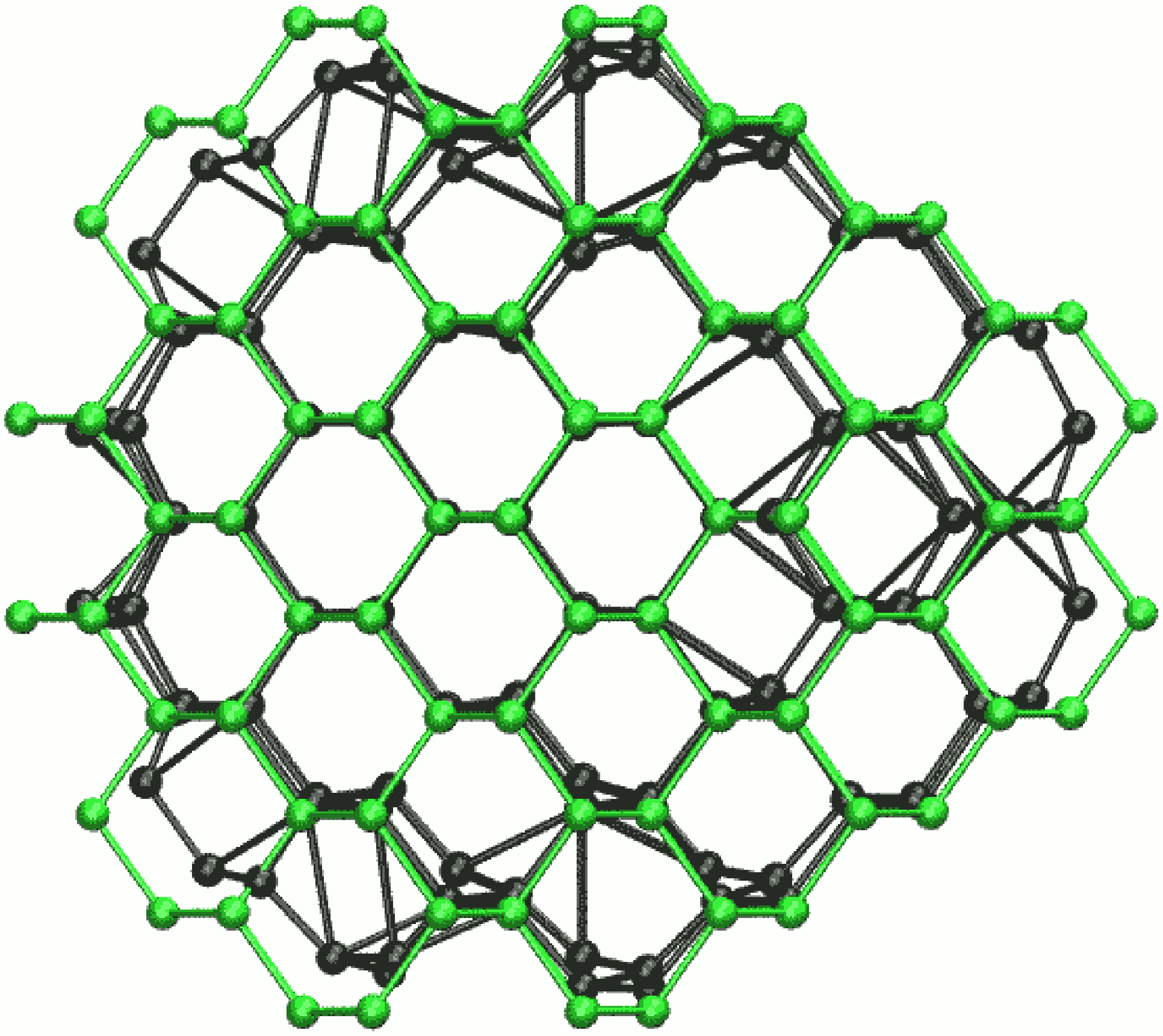}\hfill
\includegraphics[width=0.30\textwidth]{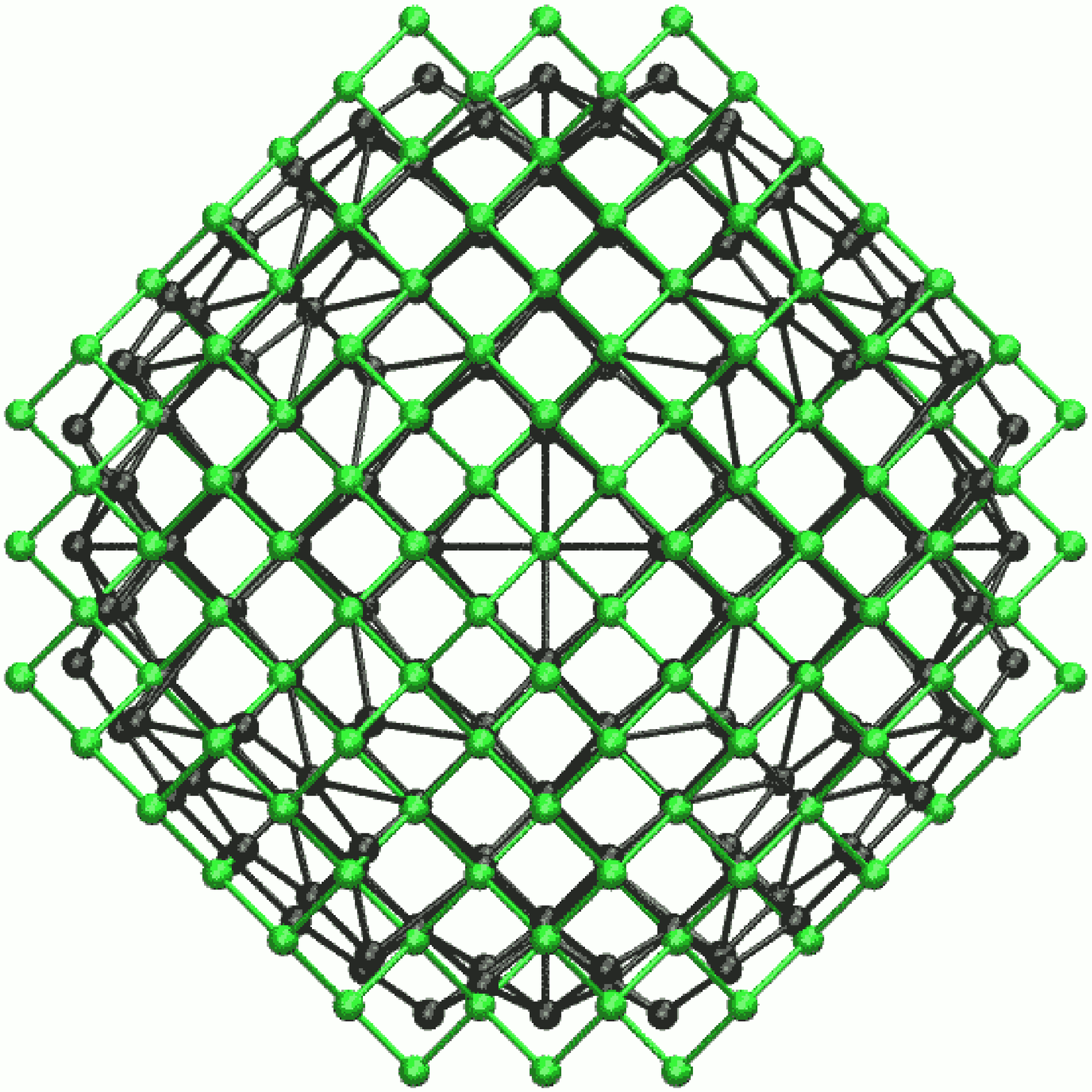}\\
\parbox[t]{\textwidth}{\caption{(\textbf{a}) Magnitude of the atom position shift to the origin vs. coordinate (fitted to the function (\ref{eq:deformation}) with parameters $s = 10$, $a = 0.08$). (\textbf{b, c}) Compression of the $C_{302}$ diamond ball according to the function (\ref{eq:deformation}) in two projections: initial diamond ball is shown in green, deformed diamond ball -- in black.} \label{fig:deform} }
\end{center}
\end{figure*}

The chosen compression method (\ref{eq:deformation}) allows roughly to estimate the relaxed diamond ball shell structure under surface forces resulting from self-consistency of the electron-nuclear interaction. A quantitatively correct description of the compression is much more complex and, more importantly, would provide higher precision than we really need for demonstration purposes.

\subsection{\textit{Ab initio} diamondoids electronic structure computations}

Ground-state electronic structure computations for several small diamondoids $C_{78}H_{64}$, $C_{123}H_{100}$ and $C_{211}H_{140}$ were performed at DFT~R-B3LYP level of theory using the 6-31G(d,p) basis set. Three cases were considered: non-deformed particle and two deformed configurations with compression parameters $s = 10$, $a = 0.04$ and $s = 10$, $a = 0.08$ applied according to eq.~(\ref{eq:deformation}). Computations were made in the GAMESS-US~\cite{Schmidt1993} software package (May~1,~2012~R2 version). Several pure RHF computations were performed with the same basis set and have shown that wavefunction morphology is preserved in both RHF and DFT computations, thus allowing one to use Kohn--Sham (KS) orbitals as a qualitatively correct basis for molecular orbital picture interpretation. Strictly speaking, KS orbitals ``are physically sound and may be expected to be more suitable for use in qualitative molecular orbital theory than either Hartree--Fock or semiempirical orbitals''~\cite{Baerends1997}, especially in the case of occupied states~\cite{Stowasser1999}.

The structure of hydrogenated diamondoids was optimized using OpenBabel~\cite{OBoyle2011} in the Avogadro~\cite{Hanwell2012} software package and MMFF94~\cite{Halgren1996} force field. Carbon positions were fixed in case of deformed diamondoids and only the hydrogen shell was relaxed.

\section{Quantum-mechanical analysis of model diamond balls}

We have found that every bonding state in diamond balls is intrinsically collective and we classify them according to three distinct types: bonding orbitals morphologically similar to hydrogenic atom wavefunctions, Tamm surface electron states and subsurface-localized electron states. The results are illustrated both by a 1D one-particle example and by the corresponding \textit{ab initio} computations.

\subsection{Bonding orbital morphologies in diamond balls}

\begin{figure*}[t]
\begin{center}
\center
\parbox[t]{0.33\textwidth}{\center \textbf{a}}\hfill
\parbox[t]{0.33\textwidth}{\center \textbf{b}}\hfill
\parbox[t]{0.33\textwidth}{\center \textbf{c}}\\
\includegraphics[width=0.33\textwidth]{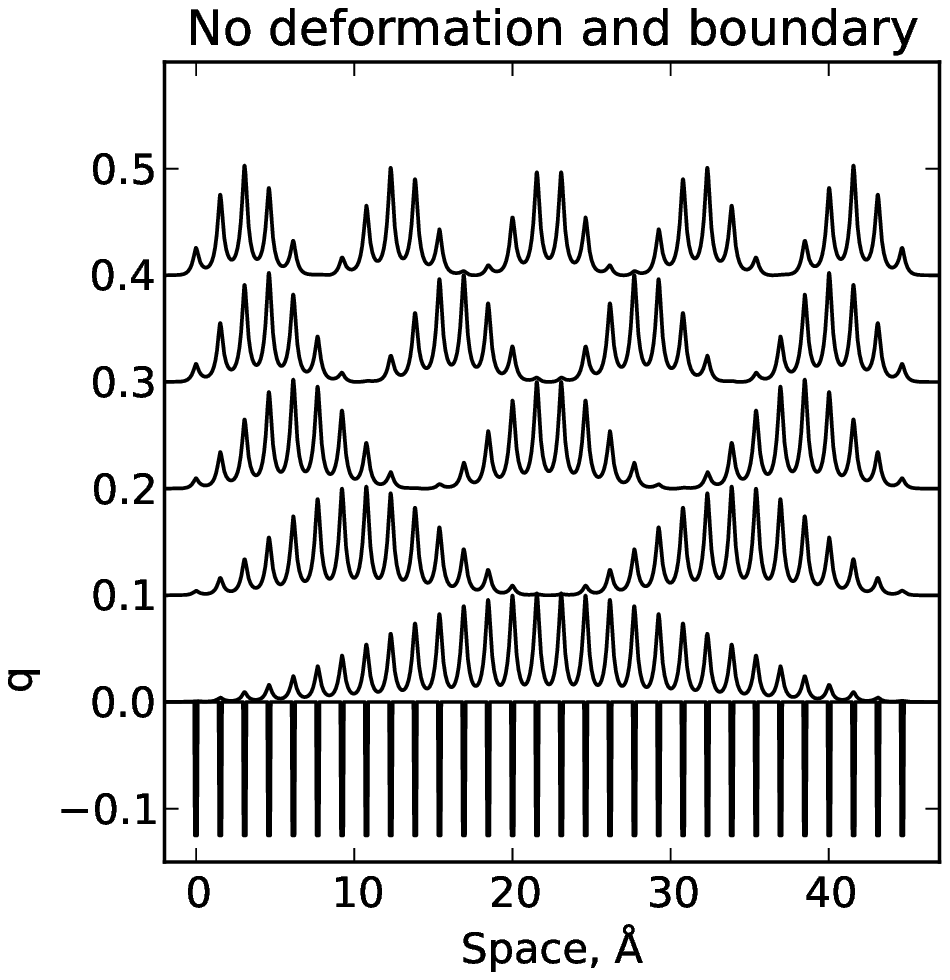}\hfill
\includegraphics[width=0.33\textwidth]{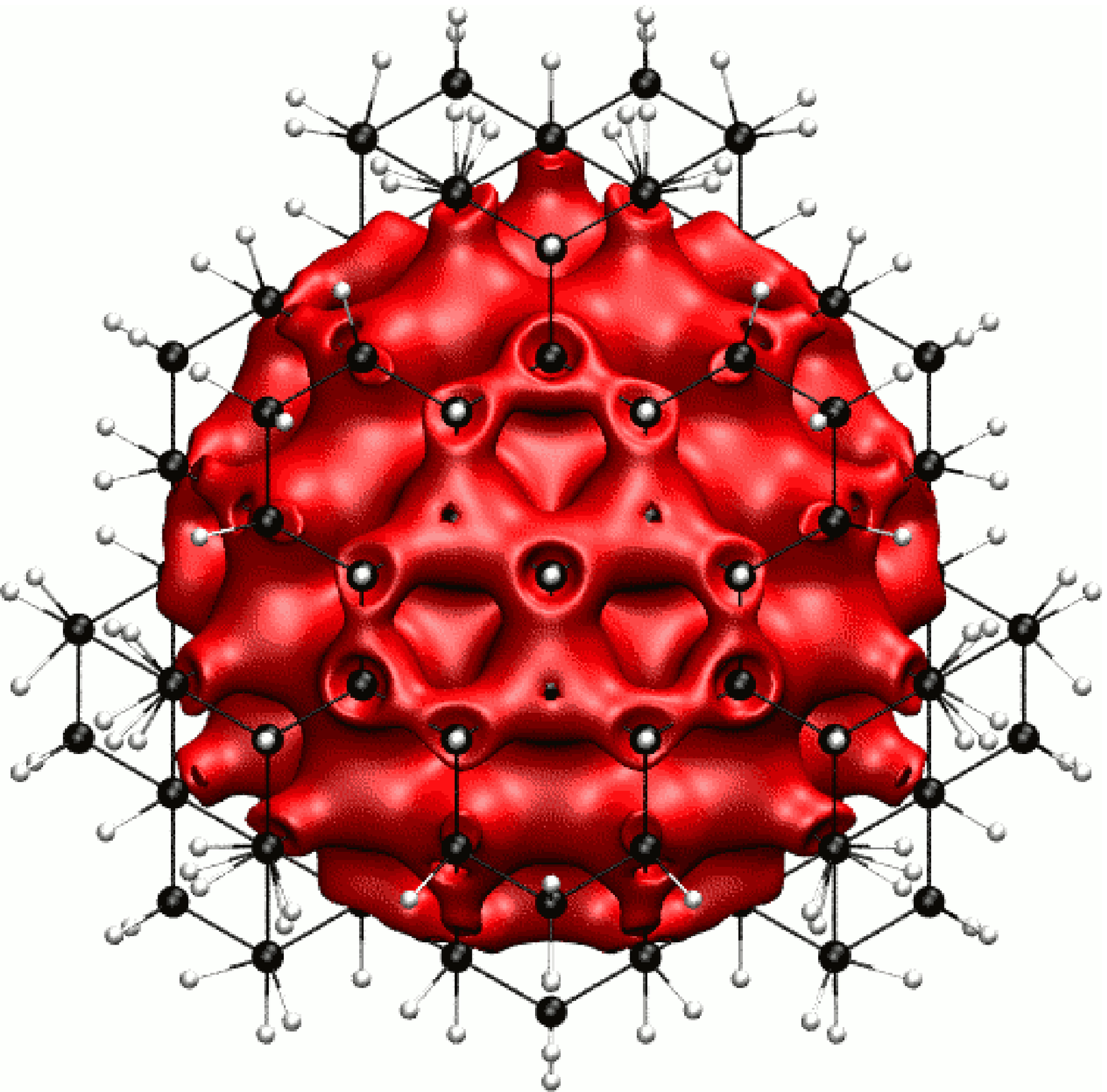}\hfill
\includegraphics[width=0.33\textwidth]{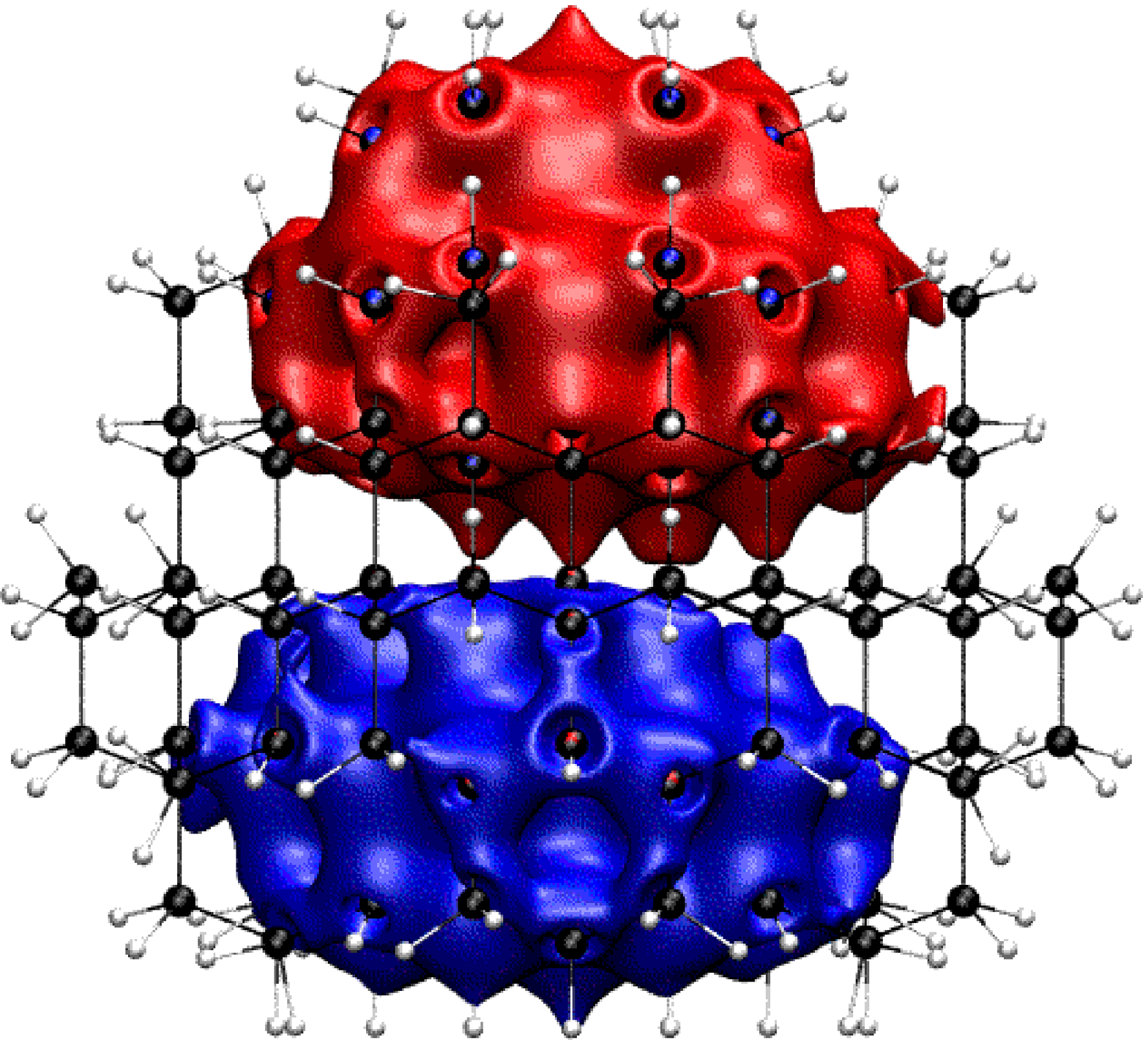}
\parbox[t]{\textwidth}{\caption{(\textbf{a}) First two bonding molecular orbitals for 1D case are resembling modulated particle-in-a-box solutions. (\textbf{b, c}) Isosurface of the first two bonding molecular orbitals \#212 and \#213 of $C_{211}H_{140}$  molecule resemble typical shapes and nodal structure of atomic s and p orbitals. Visualization is made in the VMD~\cite{Humphrey1996} software package using PovRay (\textit{http://www.povray.org/}).}
\label{fig:compare}}
\end{center}
\end{figure*}

It is worth noting that bonding molecular orbitals for 1D one-particle solutions and for spherical diamondoids (Fig.~\ref{fig:compare}) give solutions with shapes and nodal structure similar to modulated hydrogenic atom wavefunctions of s-, p-, d-shape, etc. Generally, the observed orbital morphologies arise from near-spherical symmetry of the system and similarity with the particle-in-a-box case.

\subsection{Tamm surface collective electron states}

\begin{figure*}[t]
\begin{center}
\includegraphics[width=\textwidth]{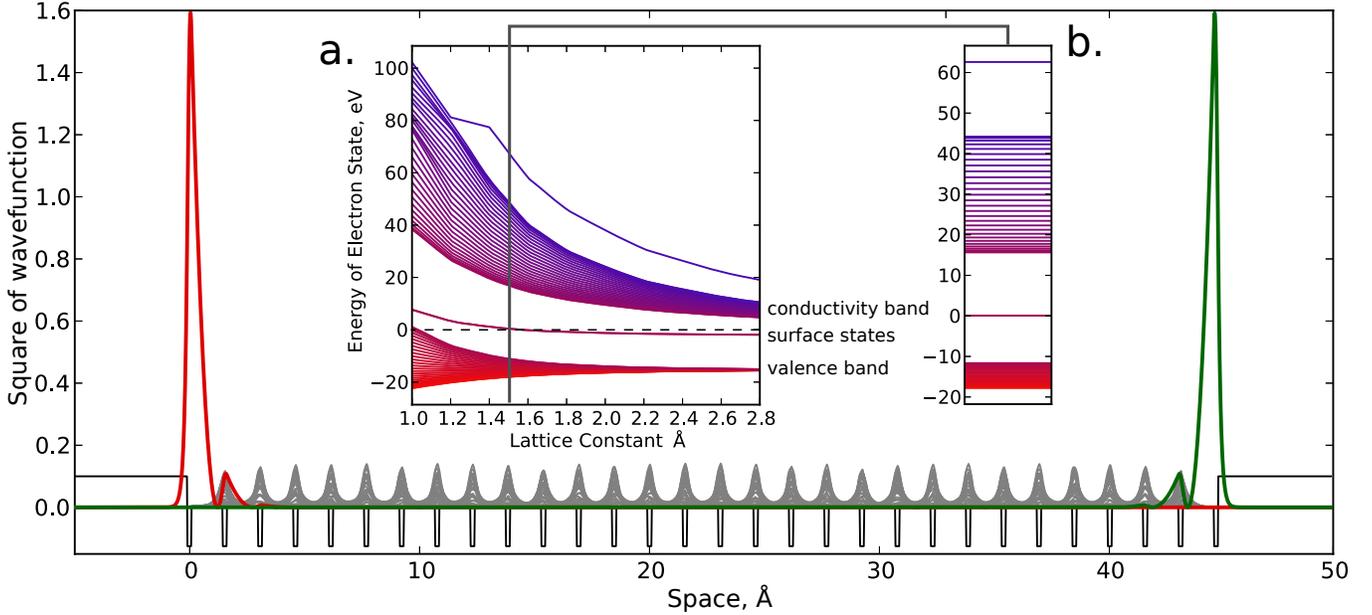}
\caption{Square of electron wavefunctions in limited periodic potential (holes deepness 25\,eV, width 0.2\,\AA, lattice constant 1.54\,\AA, boundary 5\,\AA$\times$20\,eV). (a) Energy spectrum versus lattice constant of one-dimensional crystal. (b) Energy spectrum for 1.54\,\AA~lattice constant.}
\label{fig:surfaceStates}
\end{center}
\end{figure*}

As shown in Figure~\ref{fig:surfaceStates}, the electron density of some quantum states (bold lines on the plot) is located near the surface layer and their energy levels are located between the valence and conduction bands. The total number of states in valence and surface bands are equal to the number of potential wells. In accordance with earlier discussed classical theoretical results surface state penetration into the crystal is observed if potential holes are shallow enough. The  energy level of the surface states is almost parallel in the plot of energy (Fig.~\ref{fig:surfaceStates}) versus lattice constant, and can be either below or above the Fermi energy (denoted as zero line).

\begin{figure*}[t]
	\begin{center}
	\center
	\begin{tabular}[c]{cc}
		\textbf{a}. $C_{211}H_{140}$ , $a = 0.00$, HOMO&
		\textbf{b}. $C_{211}H_{140}$ , $a = 0.08$, HOMO\\
		\includegraphics[scale=0.3]{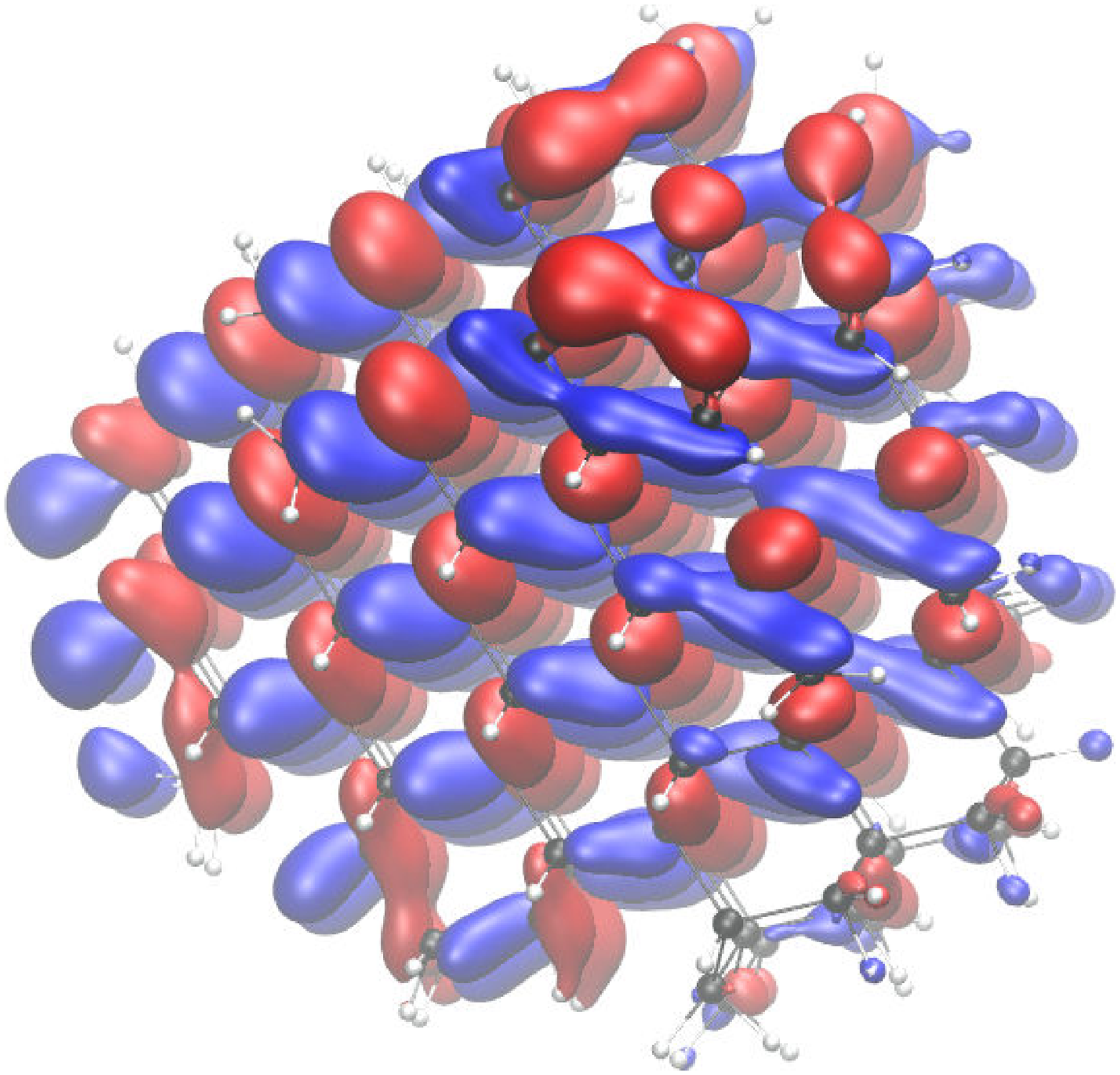}&
		\includegraphics[scale=0.3]{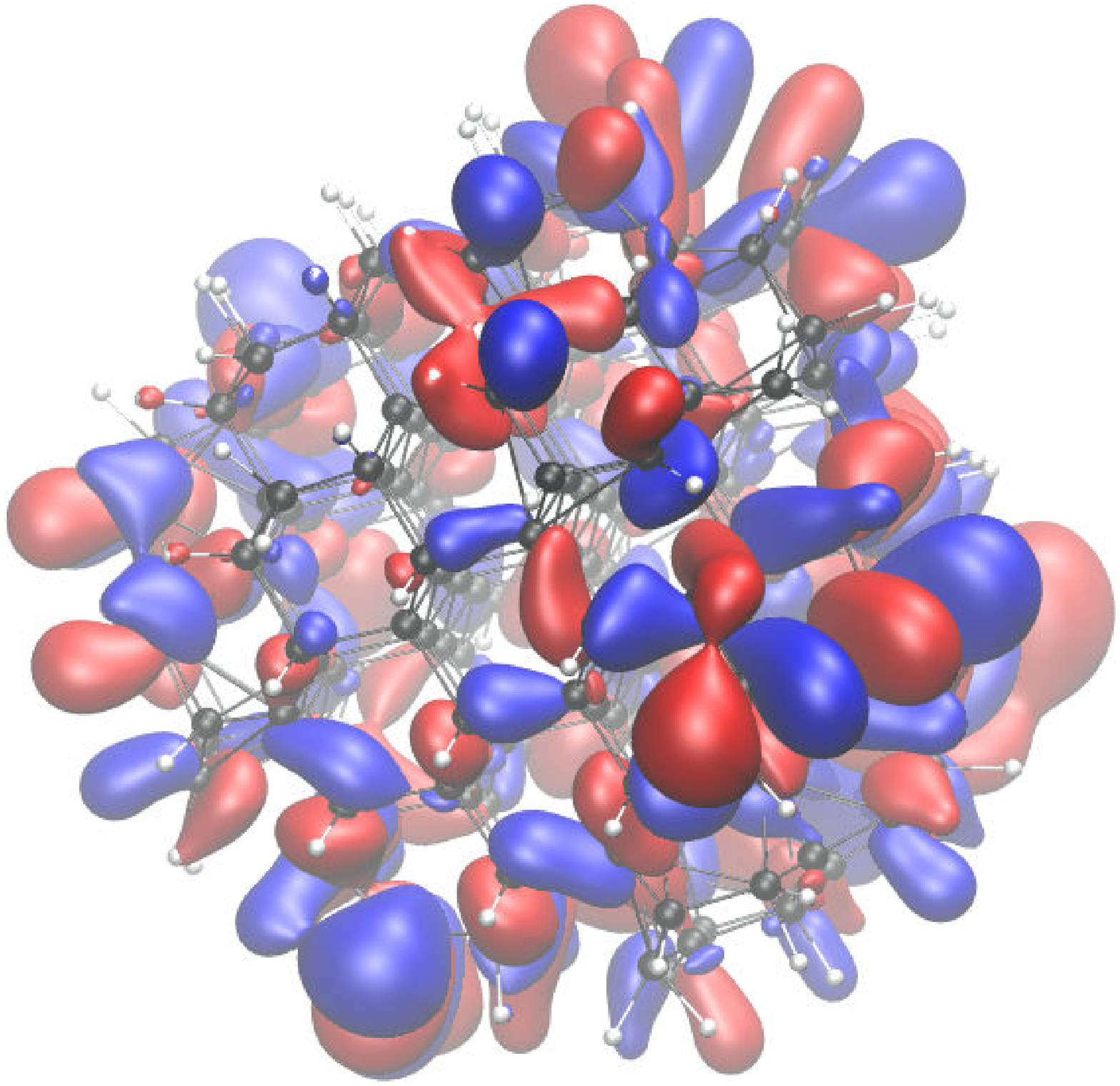}\\
		\textbf{c}. $C_{211}H_{140}$ , $a = 0.00$, LUMO& 
		\textbf{d}. $C_{211}H_{140}$ , $a = 0.08$, LUMO\\
		\includegraphics[scale=0.3]{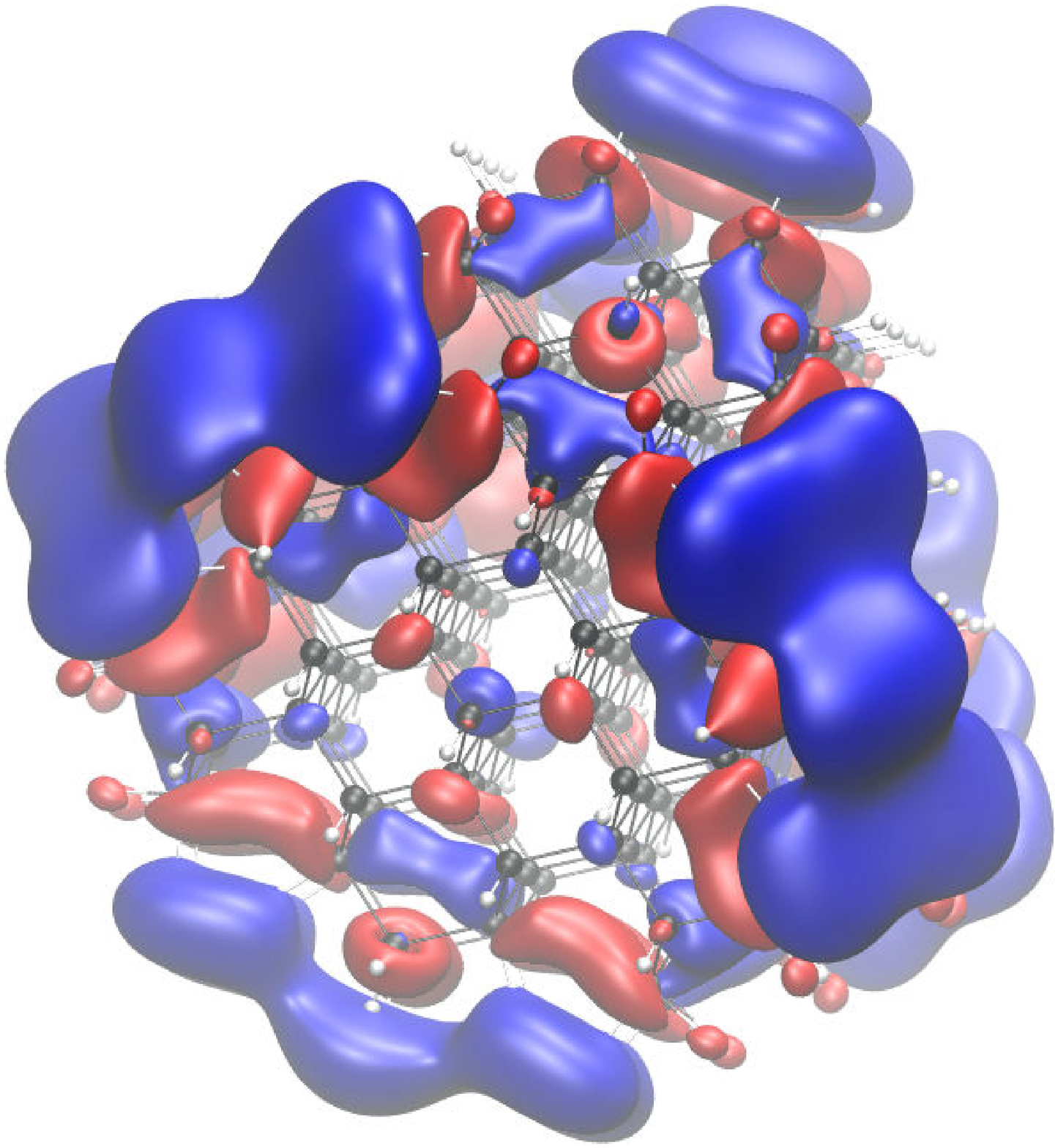}&
		\includegraphics[scale=0.3]{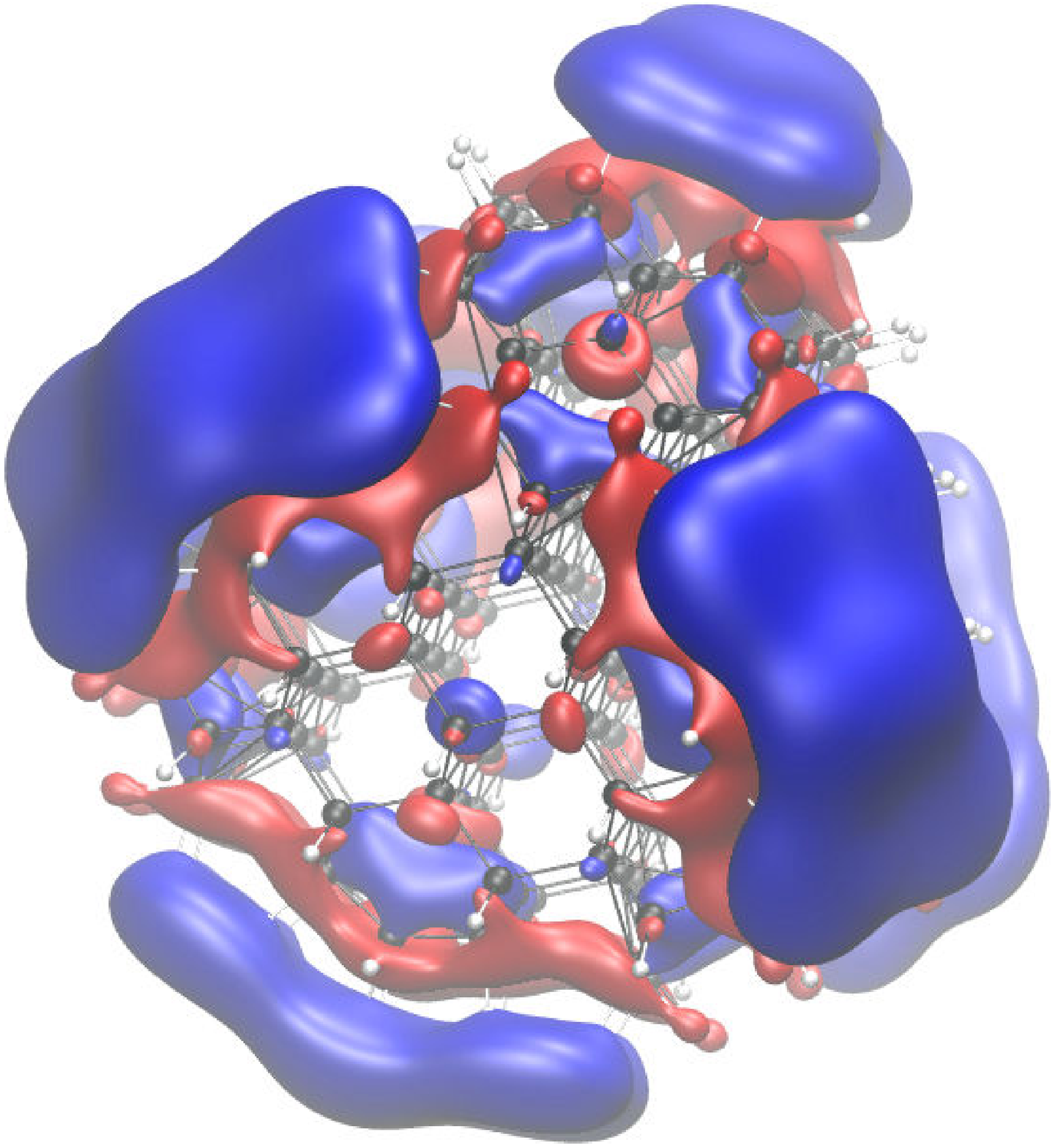}
	\end{tabular}
	\parbox[t]{\textwidth}{\caption{Isosurfaces of wavefunction of 1.34\,nm diamondoid $C_{211}H_{140}$ (red $0.01$\,a.u., blue $-0.01$\,a.u.): (\textbf{a}) HOMO, $a = 0.00$, (\textbf{b}) HOMO, $a = 0.08$, (\textbf{c}) LUMO, $a = 0.00$, (\textbf{d}) LUMO, $a = 0.08$. } \label{fig:surface3d}}
	\end{center}
\end{figure*}

Consideration of the 3D problem allows to investigate spatial localization of Tamm states. It's unclear from 1D calculations if Tamm states are local in a sense of dangling bond localization or have intrinsically collective nature. It's still impossible to compute the optimized electronic structure of real 5\,nm diamond ball in the three dimensional case. That is why we have investigated surface states of computationally feasible 1.34~nm diamondoid $C_{211}H_{140}$. Tamm states in the 3D singlet case are located inside virtual orbitals band and show collective and surface-localized wavefunction nature. The HOMO and LUMO states for the compressed and non-compressed diamondoid $C_{211}H_{140}$ are shown on Fig.\,\ref{fig:surface3d}.

\subsection{Subsurface collective electron states and compression factor}

In this section the influence of compression on diamondoid wavefunction morphologies is investigated. Assuming that 70\% of diamond ball $sp^3$-bonds are deformed~\cite{Fang2009}, the radius of the uncompressed core of the diamond ball approximately equals to $2/5$ of the ball radius.

\begin{figure*}[t]
\begin{center}
\includegraphics[width=\textwidth]{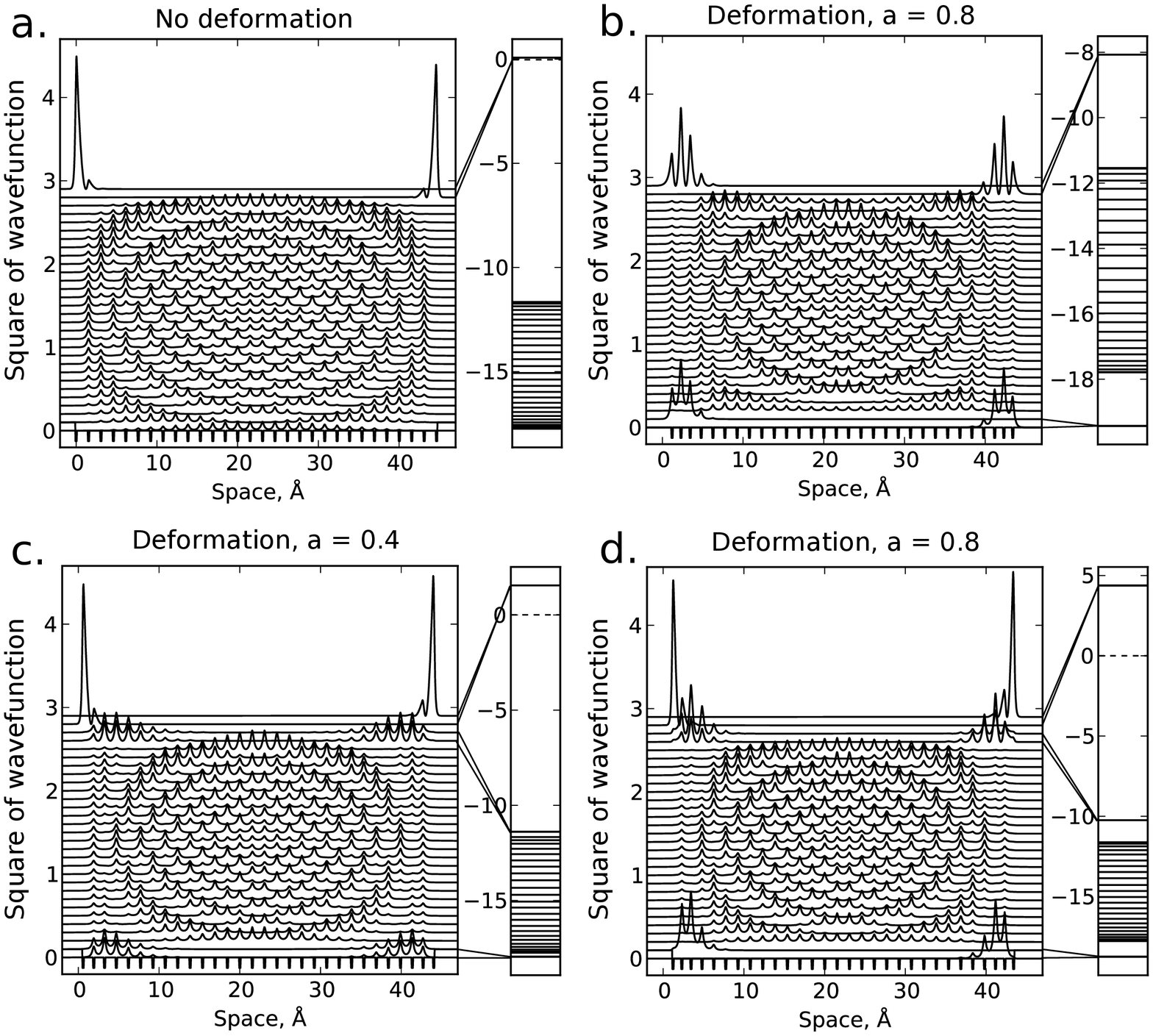}
\caption{Collective 1D electron states for: (\textbf{a}) Tamm surface states with barrier, (\textbf{b}) subsurface states in presence of compression, but without barrier, (\textbf{c}) subsurface states in presence of barrier and weak compression, (\textbf{d}) subsurface states in presence of barrier and strong compression.}
\label{fig:deform1D}
\end{center}
\end{figure*}

As can be seen from Fig.~\ref{fig:deform1D},\,a surface Tamm states are localized on both sides of crystal. Symmetrical shift of the edge potential holes according to (\ref{eq:deformation}) results in subsurface localization of wavefunctions (Fig. \ref{fig:deform1D},\,b). Localization change is observed in the lowest and highest (nearest to the surface states) wavefunctions. Subsurface states show a symmetrical split in the case of relatively low compression, becoming asymmetrical in the presence of high compression and show significant splitting from the valence band (Fig. \ref{fig:deform1D},\,b).

\begin{figure*}[t]
	\begin{center}
	\center
	\begin{tabular}[c]{p{0em} c}
	\raisebox{-0.3\height}{\includegraphics[scale=0.6]{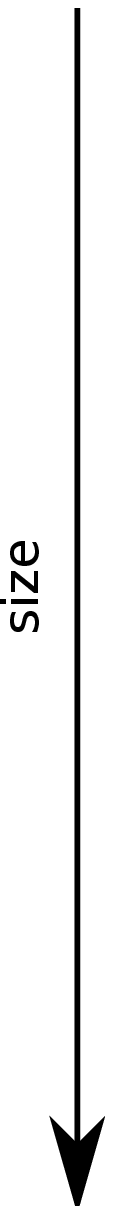}} & 
	\begin{tabular}[c]{ccc}
		\multicolumn{3}{c}{\includegraphics[scale=0.6]{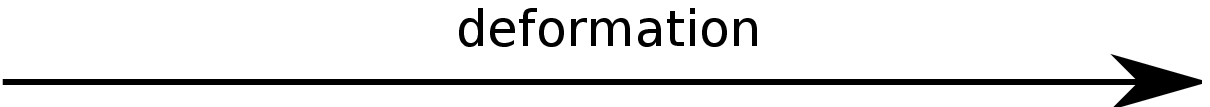}} \\
		\textbf{a}. $C_{78}H_{64}$, $a = 0.00$& 
		\textbf{b}. $C_{78}H_{64}$, $a = 0.04$&
		\textbf{c}. $C_{78}H_{64}$, $a = 0.08$\\
		\includegraphics[scale=0.25]{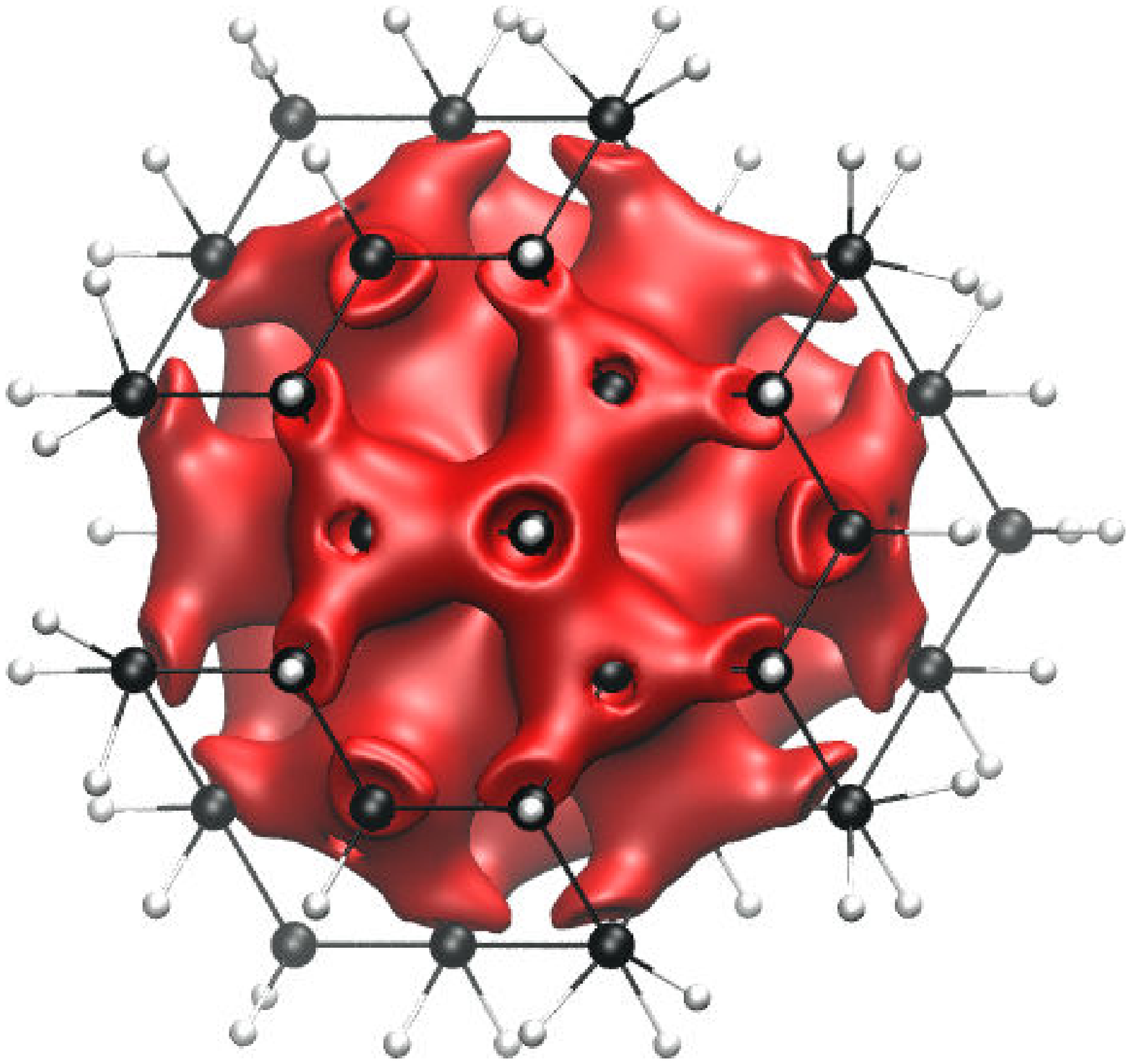}&
		\includegraphics[scale=0.25]{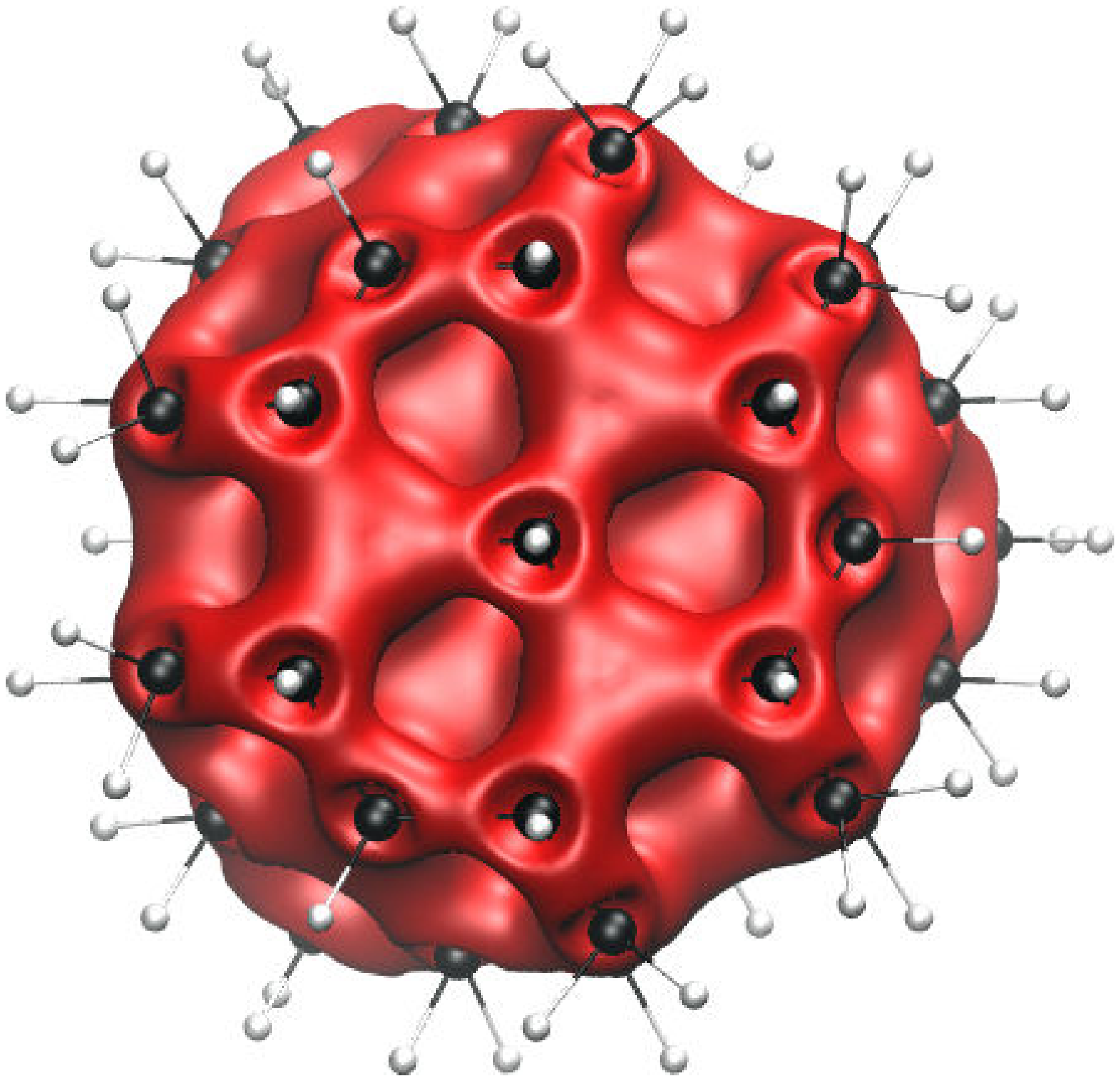}&
	    \includegraphics[scale=0.25]{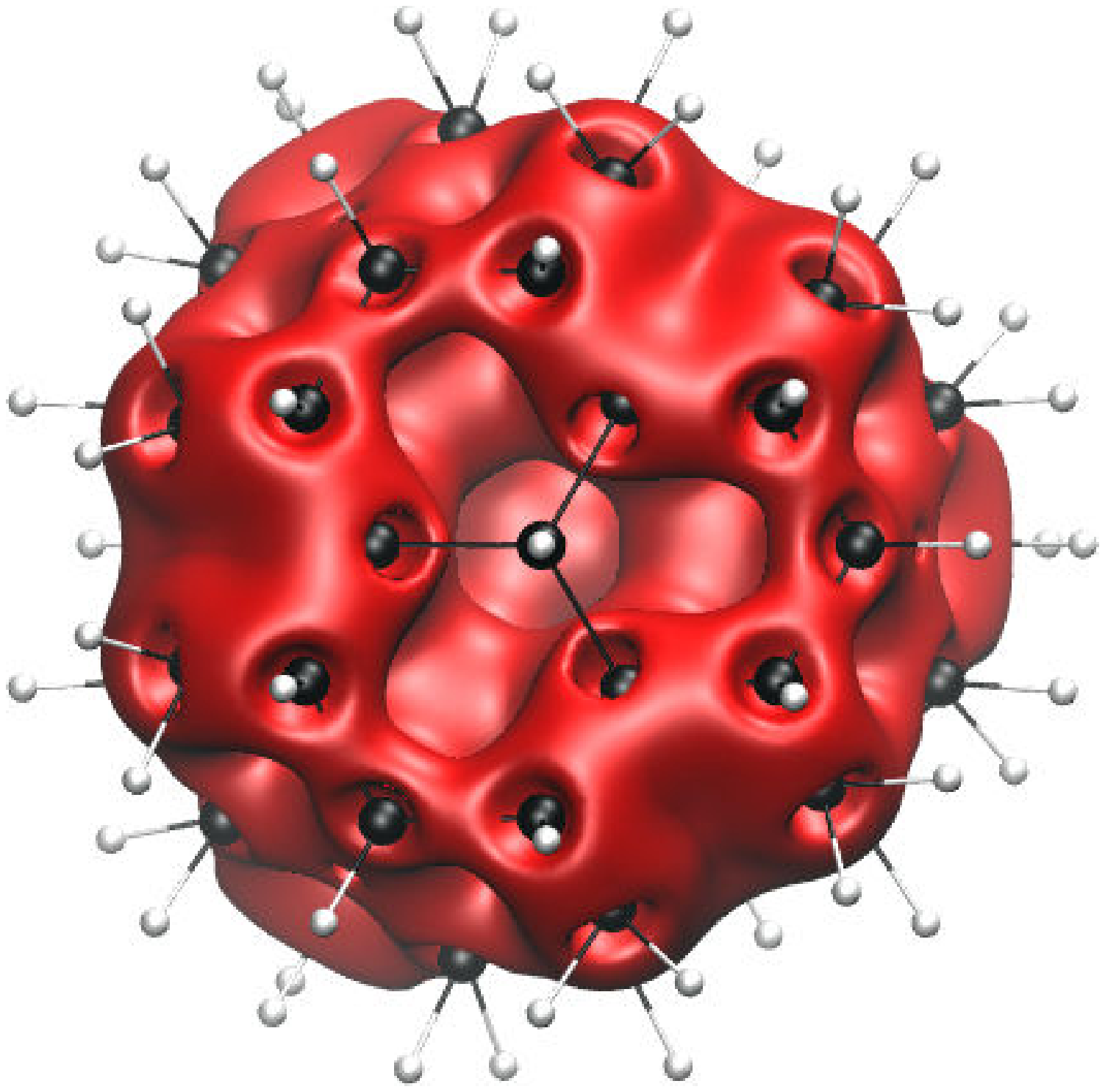}\\
		\textbf{d}. $C_{123}H_{100}$, $a = 0.00$&
		\textbf{e}. $C_{123}H_{100}$, $a = 0.04$&
		\textbf{f}. $C_{123}H_{100}$, $a = 0.08$\\
		\includegraphics[scale=0.25]{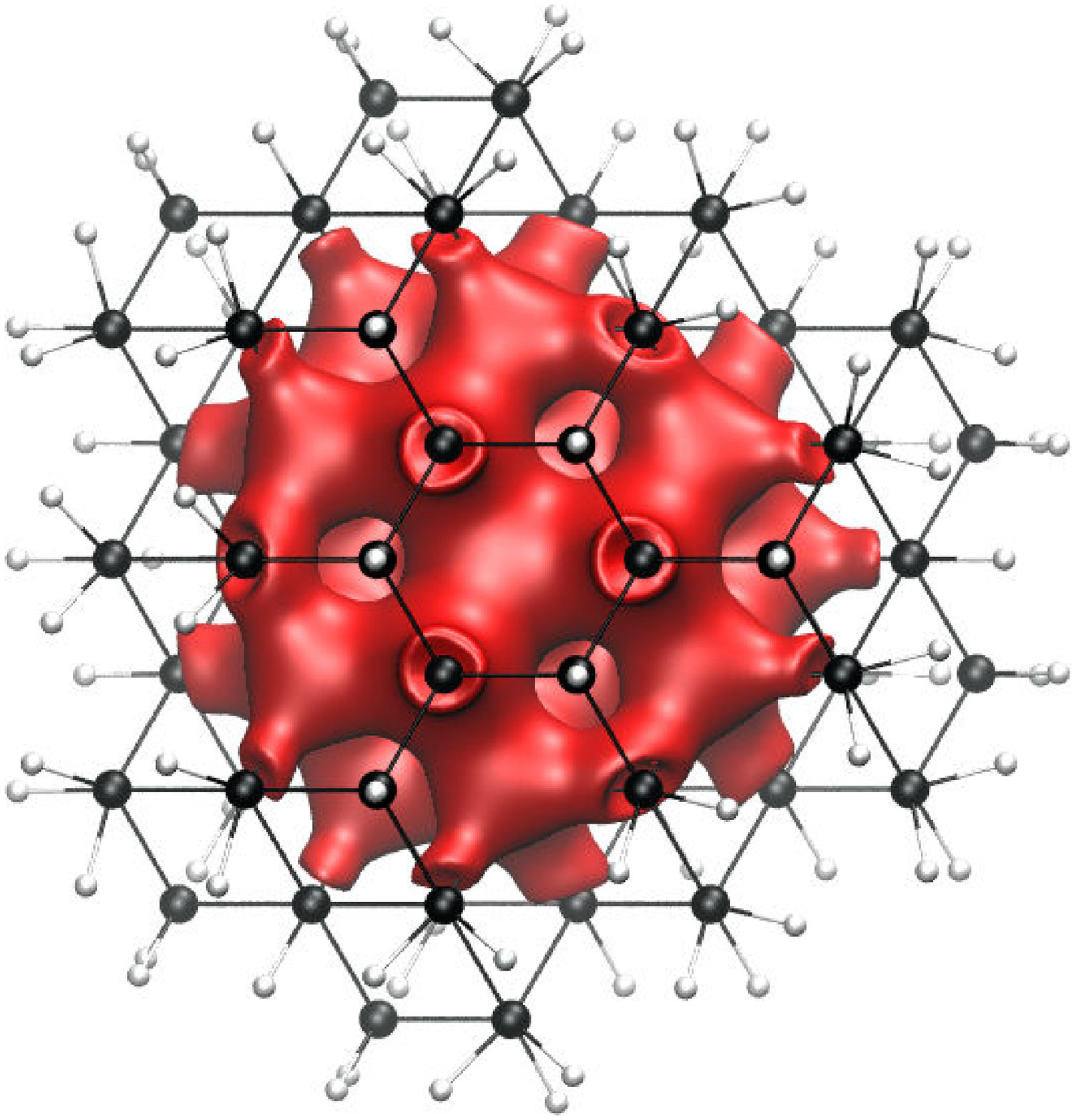}&
		\includegraphics[scale=0.25]{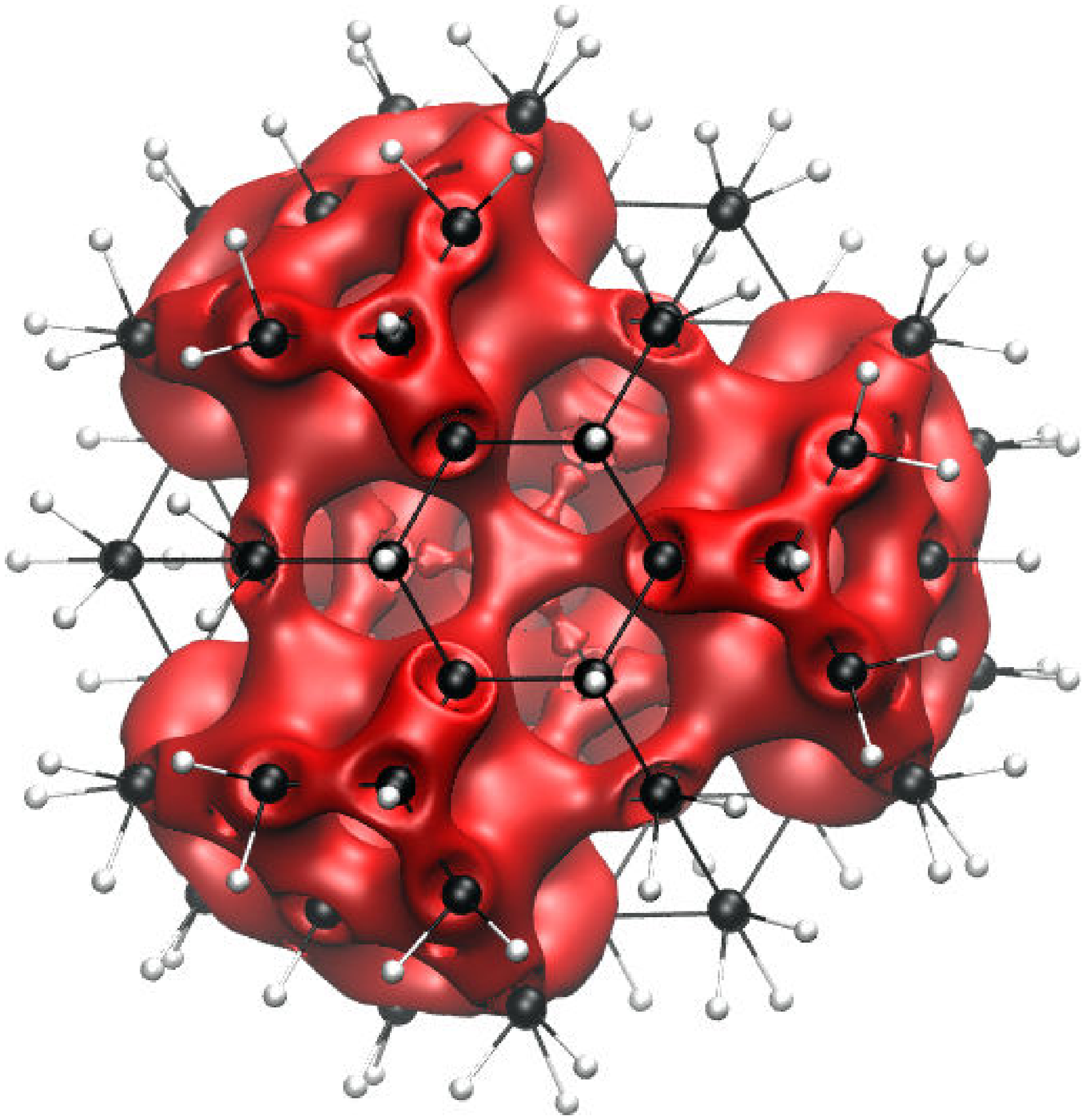}&
		\includegraphics[scale=0.25]{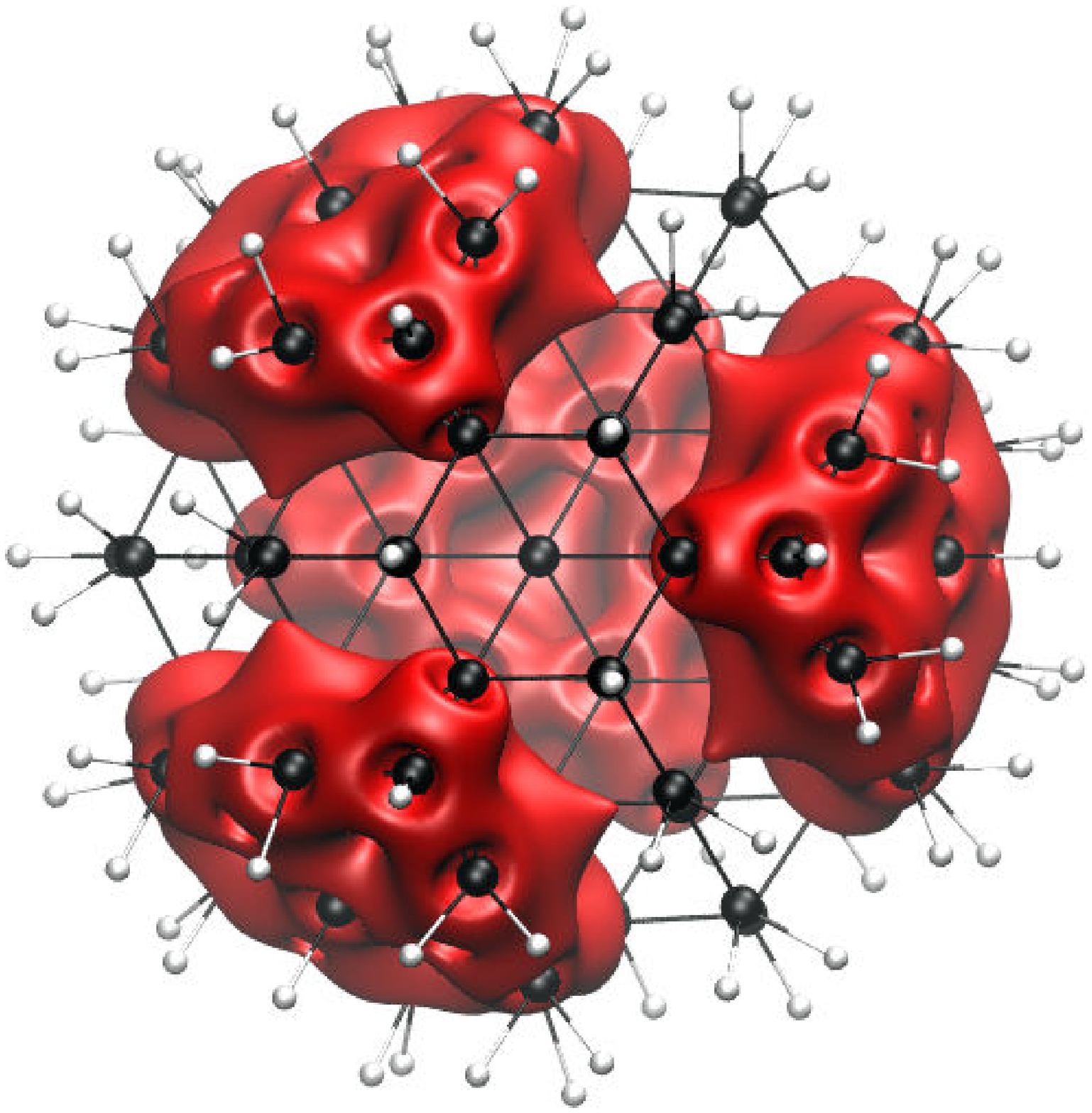}\\
		\textbf{g}. $C_{211}H_{140}$, $s = 0.00$&
		\textbf{h}. $C_{211}H_{140}$, $s = 0.04$&
		\textbf{i}. $C_{211}H_{140}$, $s = 0.08$\\
		\includegraphics[scale=0.25]{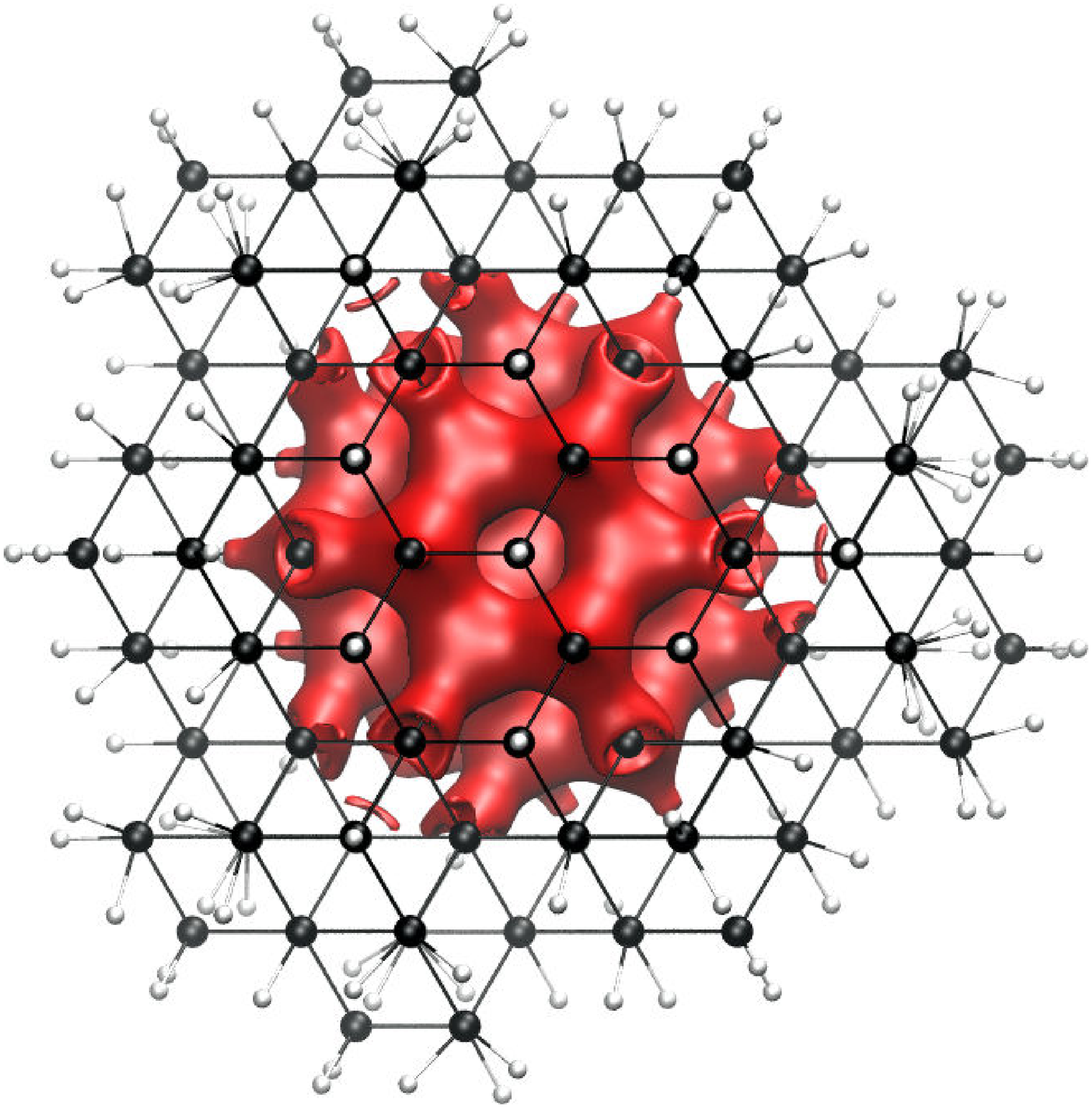}&
		\includegraphics[scale=0.25]{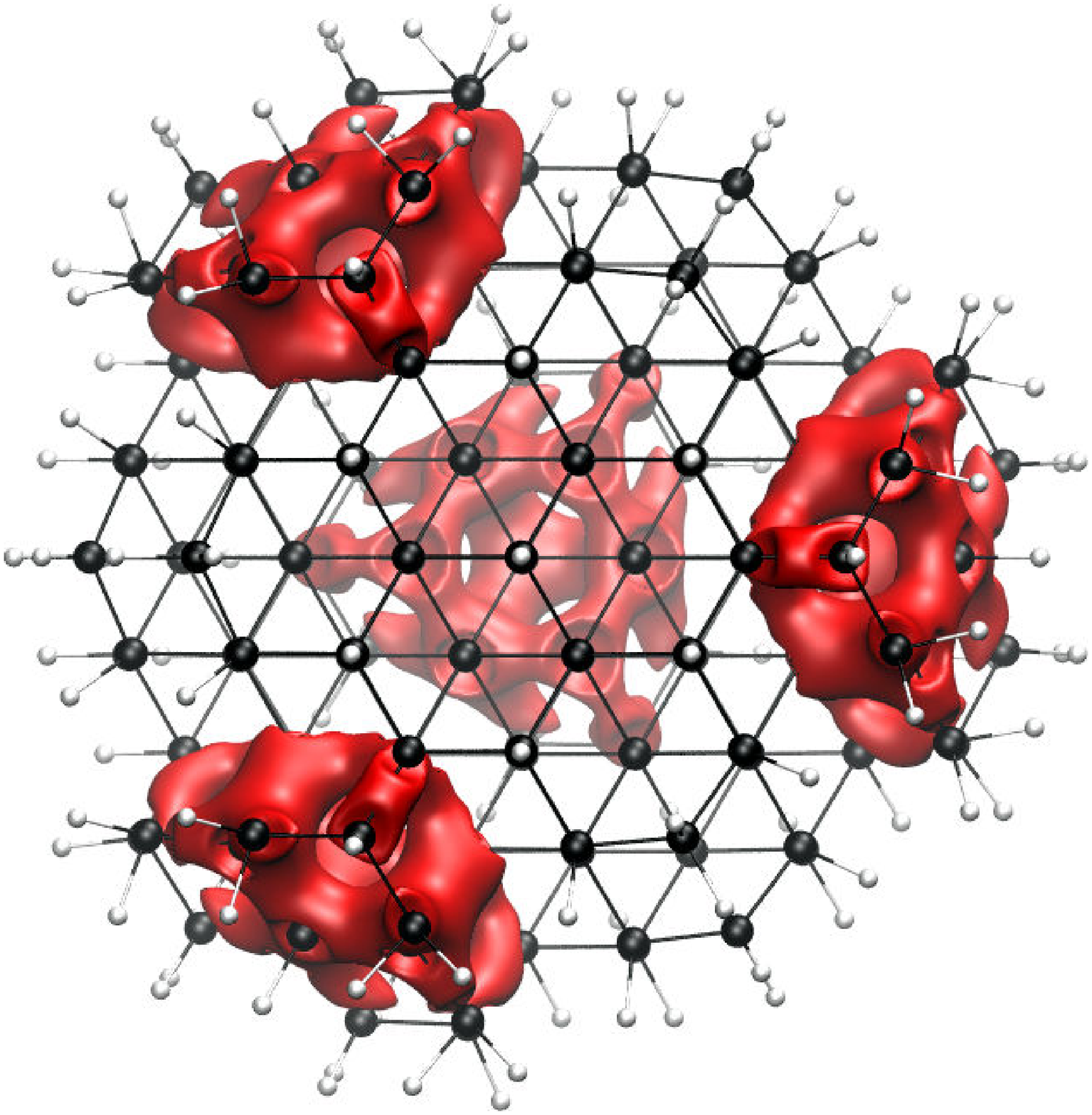}&
		\includegraphics[scale=0.25]{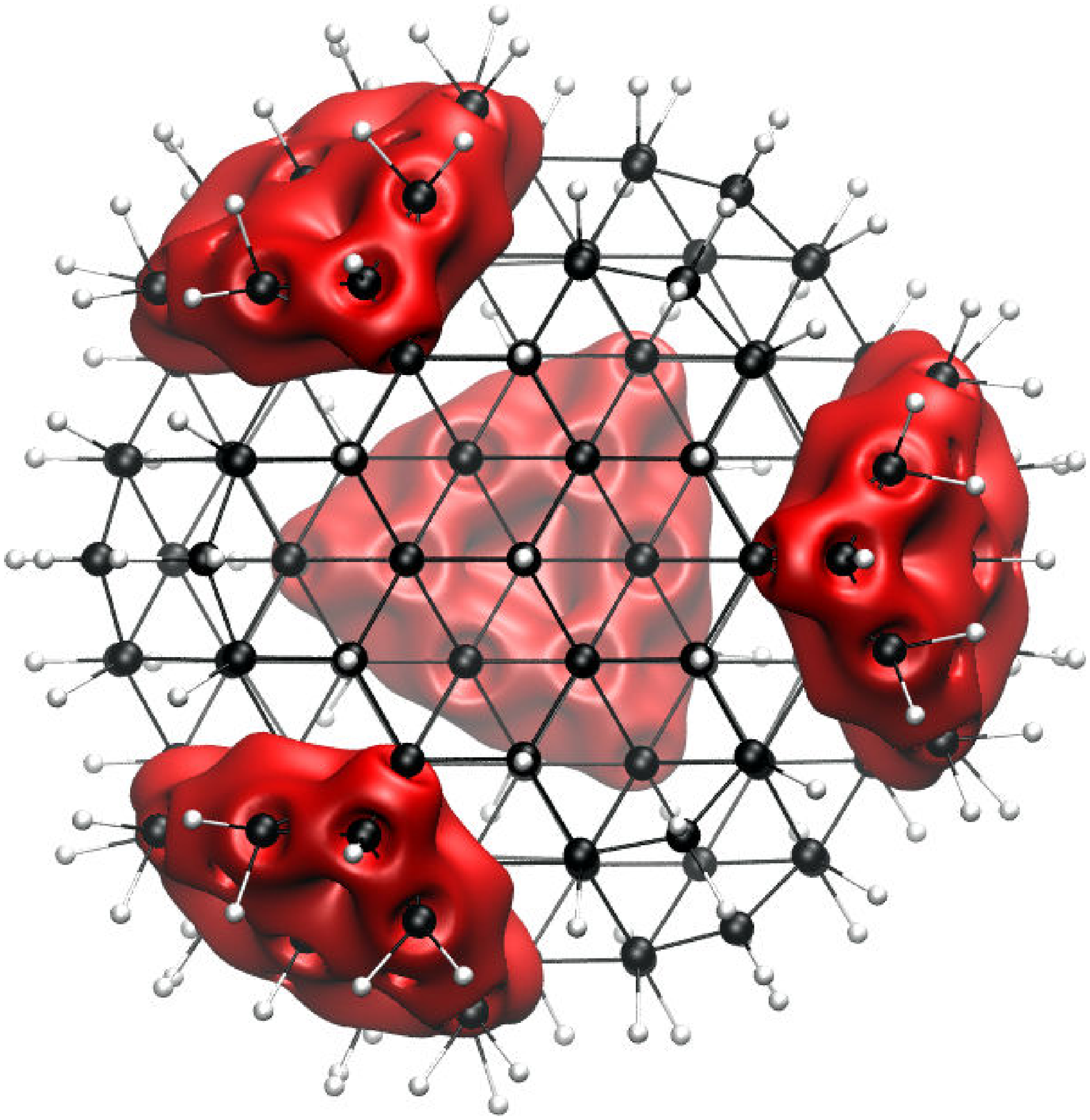}\\
	\end{tabular}
\end{tabular}
	\parbox[t]{\textwidth}{\caption{Wavefunctions isosurfaces (0.02\,a.u.) for the lowest bonding orbital of diamond balls of three sizes: (\textbf{a--c})~$C_{78}$, (\textbf{d--f})~$C_{123}$, (\textbf{g--i})~$C_{211}$ and three fixed compressions.} \label{fig:deform3D}}
	\end{center}
\end{figure*}

The lowest bonding molecular orbital morphology for uncompressed, slightly compressed and heavily compressed cases is shown in Fig.~\ref{fig:deform3D}. Nine diamondoids are considered. The general tendency could be seen in all cases, corresponding to orbital localization flowing into the subsurface area. Consideration of the $C_{78}H_{64}$ diamondoid (Fig.~\ref{fig:deform3D},\,a--c) shows slightly pronounced shift of the first bonding molecular orbital to the subsurface area, while bigger diamondoids $C_{123}H_{100}$~(Fig.~\ref{fig:deform3D},\,d--f) and $C_{211}H_{140}$~(Fig.~\ref{fig:deform3D},\,g--i) show dramatic orbital localization shift. Another interesting point is that in uncompressed diamondoids (Fig~\ref{fig:deform3D},\,a,\,d,\,g) first bonding molecular orbitals is almost identical.

\begin{figure}[t]
\begin{center}
\includegraphics[width=\linewidth]{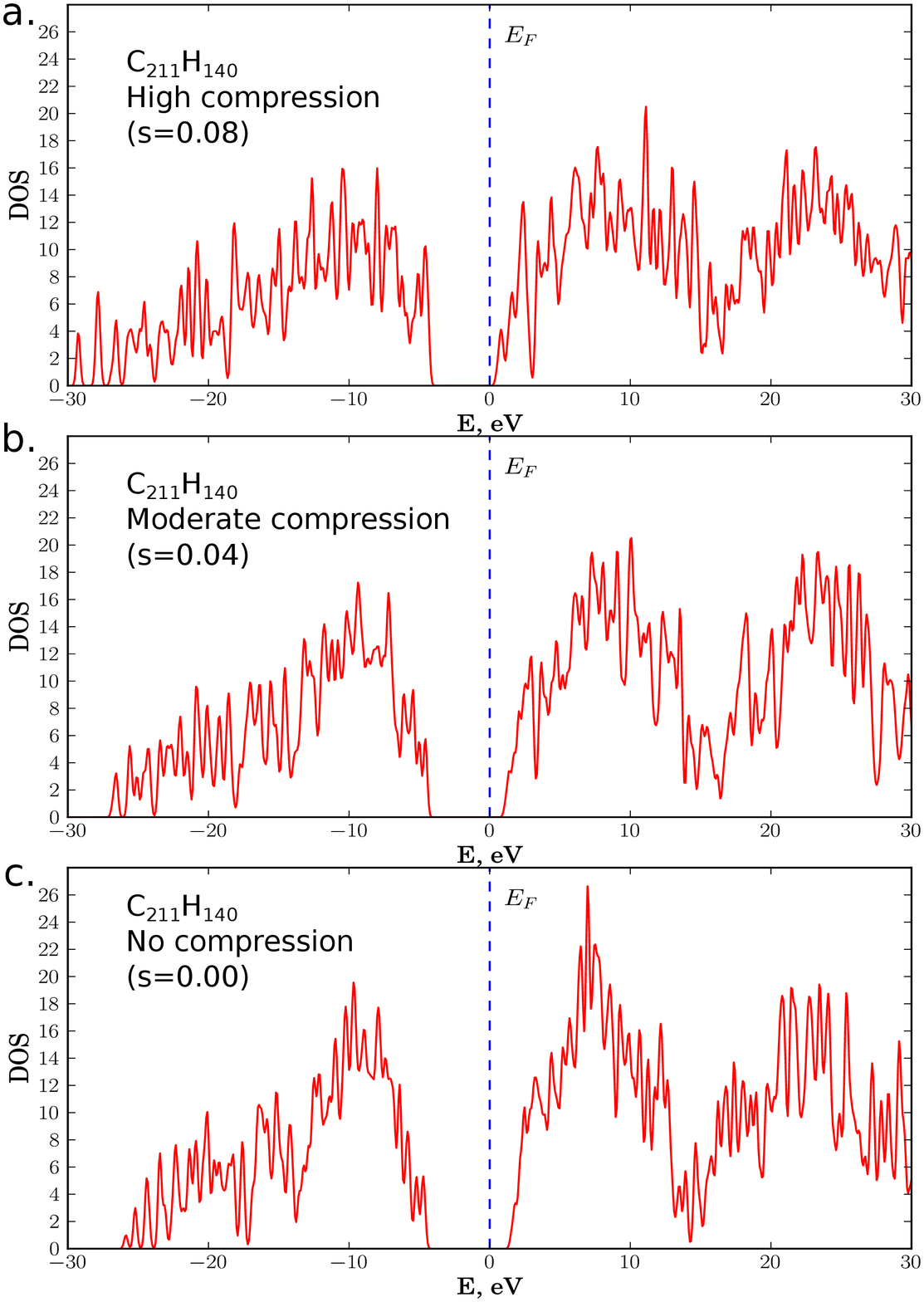}
\caption{Density of states for the $C_{211}H_{140}$ diamondoid series as a function of applied compression. Valence band bottom tend to be more populated and shift down in energy as compression is increased.}
\label{fig:dos}
\end{center}
\end{figure}

Density of states plot for the $C_{211}H_{140}$ diamondoid series is presented at the Fig.~\ref{fig:dos}. It follows that as compression is applied valence band bottom tends to shift downwards in energy, and population of the subsurface-localized states also increases. 

\section{Discussion}

Examination of our computations shows the existence of three distinct types of solutions for both 1D and \textit{ab initio} 3D cases. In this section we provide a qualitative interpretation of the nature of nanodiamond ``disordered'' shell, the EPR signal invariant and the possible source of the unusual pre-peak in PEELS experiment. Explanation is provided in terms of collective electron state localization in nanodiamonds.

\subsection{Tamm surface states}

According to Tamm reasoning for a ball-shaped dielectric nanocrystal electron s can be confined in a spherical layer between the vacuum and the periodical crystal potential. The surface electron band lies between the valence band and the conduction band. Wavefunctions of the surface states are localized near the surface and form a spherical spatial layer if we extrapolate one-dimensional case to three dimensions. Tamm electrons are locked in the radial degree of freedom but free in the angular degree of freedom, moving on the surface and belong to the whole particle as if it is a single sheet. Surface localization of wavefunctions give few solutions localized on the nanodiamond; rather they float as the surface electrons.

Such ``floating electrons'' in a bulk diamond are discussed as alternative to classical electrical contacts or leads in new electronic devices \cite{Ray2011} and become the base for diamond electronic devices. As shown in experiments, surface conductivity is sensible to modification of surface, and its existence demands hydrogen terminated surface~\cite{Maier2000}. However, unpaired electrons lie under the surface of nanocrystal according to NMR data \cite{Fang2009}, and EPR properties are also independent on surface modification \cite{Belobrov2001}. That is why Tamm surface states can not explain paramagnetic properties of nanodiamond \cite{Levin2008}. Nevertheless, we suggest to take into account surface states for future exploration of the transport properties of composite materials based on nanodiamond \cite{Gordeev2013}.

Energy localization of these states in the middle of the band gap could be easily explained if we realize that one-particle one-dimensional model represents one-dimensional doublet radical state. Semiempirical simulations of $n$-mantane ($C_{60}H_{60}$) radical electron structure~\cite{Belobrov2003} show direct correspondence with the one-dimensional case, where radical energy level lie in the middle of the band gap. If we consider singlet case, Tamm states would be located in the virtual orbitals band.

\subsection{Subsurface states}

Surface layer compression in nanodiamond is experimentally proven by PEELS data showing rapid decrease of compressed layer fraction for nanodiamonds and indicating presence of both ``perfect'' core and ``deformed'' shell layer~\cite{Bursill2001, Peng2001}. 

\begin{figure*}[t]
\begin{center}
\center
\parbox[t]{0.59\textwidth}{\center \textbf{a}}\hfill
\parbox[t]{0.39\textwidth}{\center \textbf{b}}\\
\includegraphics[width=0.575\textwidth]{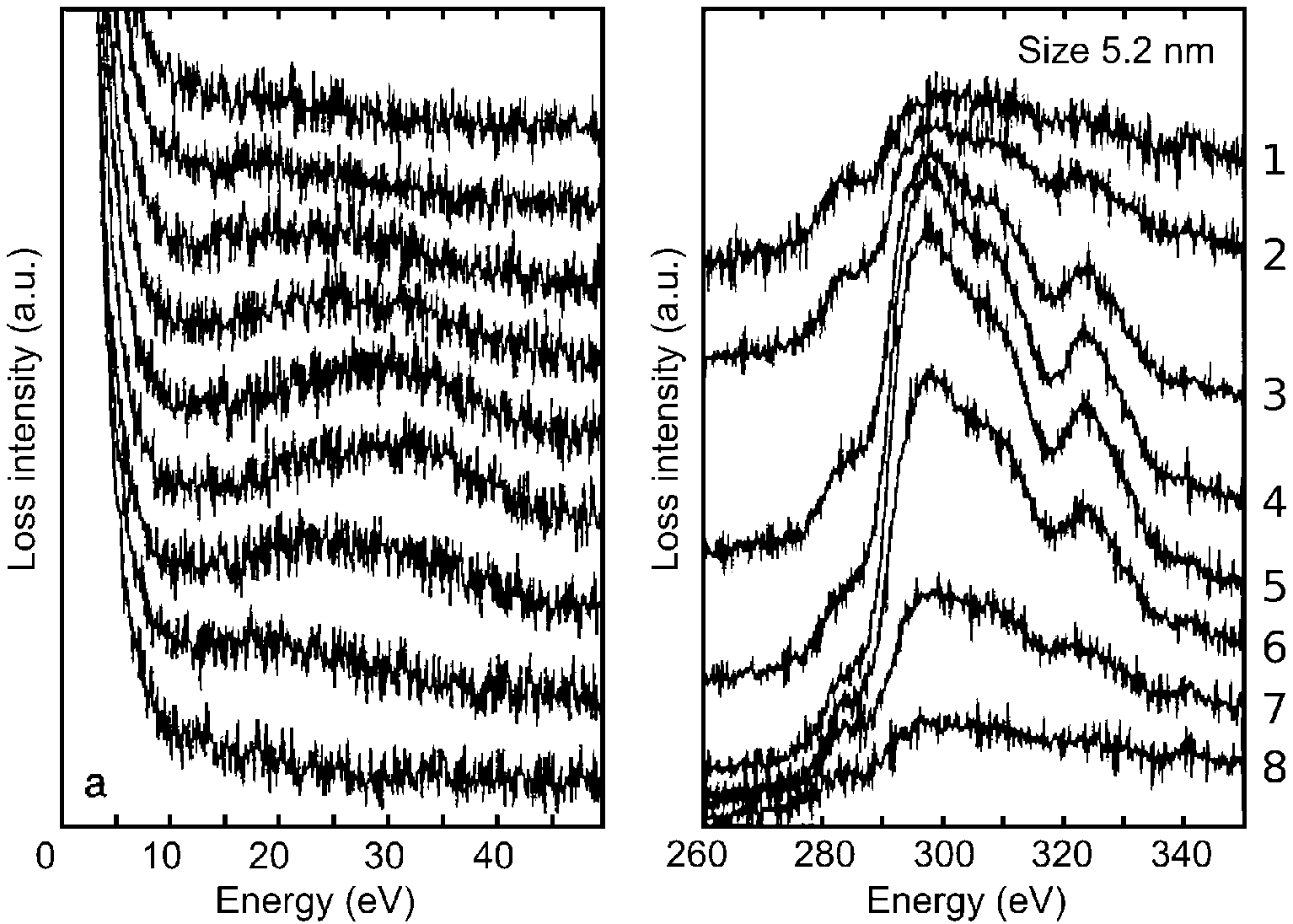}\hfill
\raisebox{0.16\height}{\includegraphics[width=0.35\textwidth]{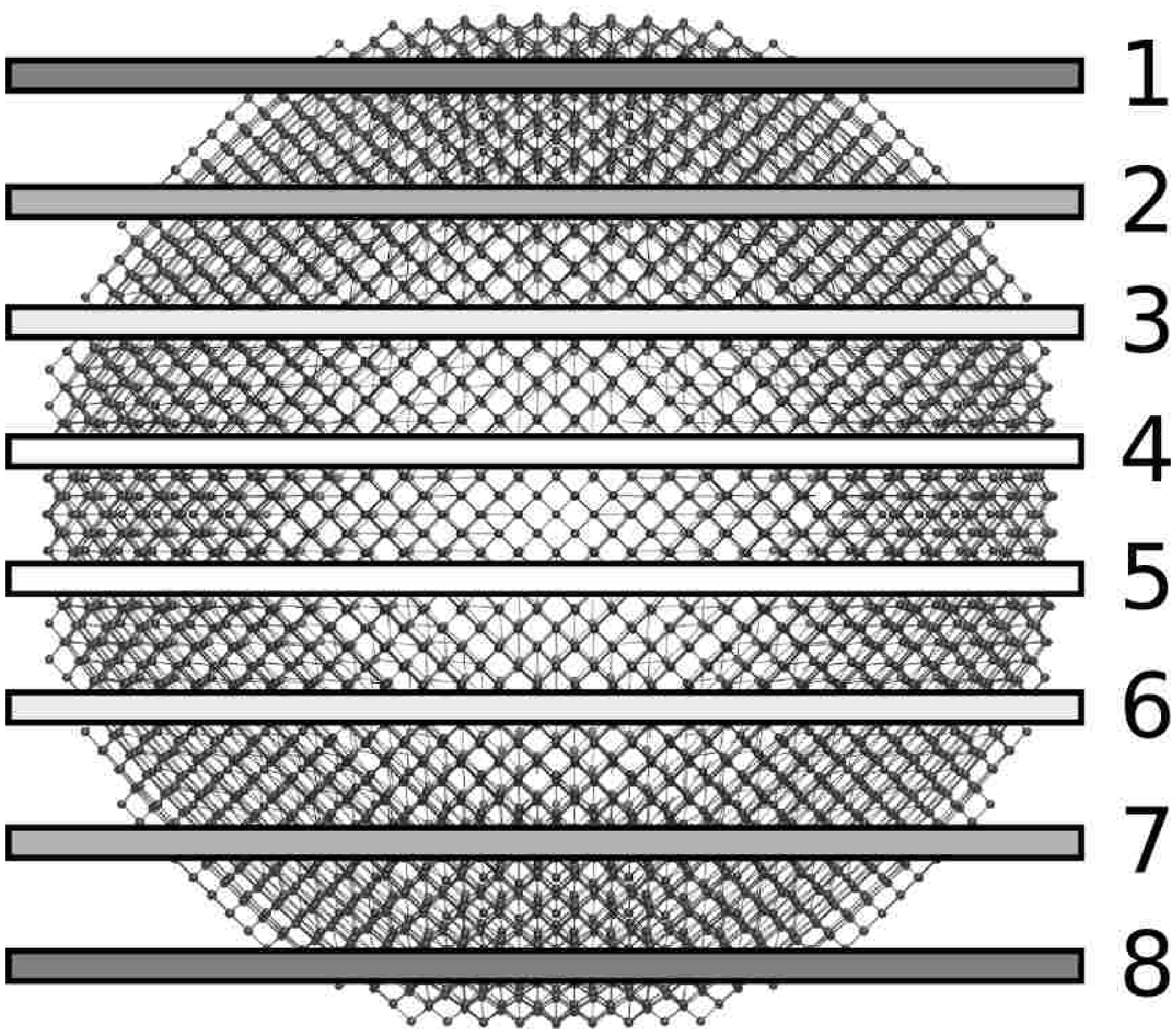}}
\parbox[t]{\textwidth}{\caption{(\textbf{a}) PEELS data for 5.2\,nm nanodiamond \cite{Bursill2001, Peng2001}. (\textbf{b}) Schematic illustration of PEELS experiment scan trajectories at 5.2\,nm compressed diamond ball.}
\label{fig:comparePEELS}}
\end{center}
\end{figure*}

PEELS data for 5.2\,nm nanodiamond is presented in Fig.~\ref{fig:comparePEELS}, where distinct character of observed curves indicate presence of both ``perfect'' core and ``deformed'' shell layer.

Quantitative measurements of the core and shell contributions allows to define deformed shell form the intensity of the pre-peak $I_{pre}$ as:
\begin{equation}
	I_{pre} = I_s / (I_s + I_b),
\end{equation}
where $I_s$ is the integrated intensity over the range 280--295\,eV of the pre-edge spectrum of diamond. This is characteristic of the distorted $sp^3$-bonded environment of the atoms under the electron probe, as it passes across the ND particle, $I_b$ is the integrated $K$ core-loss intensity over the range 295--340\,eV, corresponding to the bulk of the diamond particle, which is again dependent on the beam position on the ND particle shown schematically at Fig.~\ref{fig:comparePEELS}\,b~\cite{Peng2001, Belobrov2003}.

Compressed shell on nanodiamond particles also appears in Auger electron spectroscopy~\cite{Belobrov2003}, where clearly distinguished spectrum identifies unusual chemical states of nanodiamond. Density of states plot (Fig.~\ref{fig:dos}) indicate relative increase in the local density of subsurface-localized states. What is more, the valence band and the conduction band bottom tend to shift to the lower energy as compression is increased. These two facts could clarify the peculiarity of the Auger spectra and the pre-peak nature observed in PEELS experiment.

Additional evidence for core-shell structure of nanodiamond is presented in NMR study~\cite{Fang2009} indicating 39\% atoms in core, 40\% of partially disordered five-layered shell and remaining 21\% attributed to two nearest to surface layers.

Strictly speaking, deformation is represented by surface layer compression which is the natural consequence of electron and nuclear structure self-consistency. Recent large-scale \textit{ab initio} geometry optimization of diamondoids up to 2 nm size at DFT/6-31G within LDA level of theory shows a gradual increase in subsurface-localized character of HOMO as particles grow~\cite{Jiang2010}. It means that if one perform ``honest'' full-scale geometry optimization at an acceptable theory level subsurface states arise naturally. However, optimization of these structures is extremely demanding and time consuming.

We conclude that the rough compression model we employed allows qualitatively correct results to be obtained concerning orbital localization without employing full-scale first-principles optimization procedures.

To sum up, artificial compression gives similar subsurface states as first-principles geometrical optimization. It's argued that first-principles optimization naturally results in particle non-uniform compression leading to the subsurface localization of bonding orbitals. More importantly, these subsurface-localized bonding orbitals are collective, i.e. spread over large amount of carbon atoms located in appropriate shell.

Considering our results and large-scale computations done by Jiang et al.~\cite{Jiang2010} we extrapolate subsurface orbital model to real-sized 5\,nm nanodiamond. It appears then that these orbitals have large spatial extent, allowing us to use analogy with the long conjugated $\pi$\,-electron systems and providing an opportunity to discuss the nature of intrinsic spin in nanodiamond.

\subsection{Intrinsic spin nature}

Nanodiamond exhibits unusual spin states in wide range of experiments, including EPR spectroscopy and NMR relaxation. It demonstrates uncommon to the bulk phase paramagnetic behavior.

Unpaired electrons in nanodiamond show invariance of their $g$-factor value under experimental conditions (EPR invariant)~\cite{Belobrov2001}. In nanodiamond $g=2.0027$ against $g$-factor for a free electron $g_e=2.0023$. The difference between $g$-factor values for free electron and for unpaired electron in nanodiamond is about $10^{-4}$, which is typical for organic radicals ($10^{-3}$--$10^{-4}$). This difference depends heavily on the nearest neighborhood of the unpaired electron. Radical states centered on carbon atoms have $g$-factor value below $2.0030$. If an unpaired electron is localized on a carbon atom bonded to oxygen atom, $g$-factor value lies between $2.0030$ and $2.0040$. Oxygen-centered radical state has $g$-factor value higher than $2.0040$~\cite{Tian2009}. Thus the unpaired electrons in nanodiamond have $g$-factor values in a very narrow range and these values correspond to pure carbon atoms without any impurities in their vicinity. If the EPR signal characteristics do not depend on the surface modification including chlorination, it's source clearly doesn't belong to the surface. NMR data confirm this conclusion, indicating unpaired spin localization in subsurface layer at 0.4--1.0\,nm deepness~\cite{Fang2009, Panich2012}. There is also uncertainty in exact number of unpaired spins per particle, which ranges from from 1~\cite{Belobrov2001} in EPR study to 40~\cite{Fang2009} in NMR experiment. 

Observed unpaired electron behavior in nanodiamond couldn't be explained in terms of quite typical localized radical state, e.g. as an $F$-center, because it will necessarily yield characteristic change in optical properties of the system which is not observed experimentally. If radical state is not localized on particular defect, it must be delocalized in some sense.

To summarize, experimental data shows that unpaired electrons in nanodiamond are located in a subsurface layer and are delocalized over significant amount of carbon atoms. The unpaired electron is also completely unrelated to point defects and impurities. These facts are consistent with the subsurface localization of the first several bonding orbitals in compressed diamondoids. However, these states lie deep inside valence band and are doubly occupied, leading to fully compensated total spin. Indeed, existence of stable uncompensated spin deep inside valence band in absence of external fields would leave free position for ``upper'' electrons and is impossible. Therefore it's impossible to attribute unpaired spin of nanodiamond to any type of lattice point defects without contradicting EPR invariant and NMR results. 

Taking into account subsurface localization of these unpaired spins, we attribute them to collective subsurface orbitals. These subsurface orbitals have comparatively long extent, and spin-density fluctuations could possibly explain existence of intrinsic unpaired spin in nanodiamond. Analogy may be made with long conjugated $\pi$\,-electron systems where singlet instabilities of ground-state solutions are widely known~\cite{Fukutome1968}. Fluctuational nature of intrinsic spin is probably the only suitable explanation which does not contradict experimental data, chemical reason, charge or high-spin states issues. There is no way to imagine classical radical spin\,--\,1/2 state localized in the subsurface layer and being nevertheless stable. Moreover, it seems that the number of intrinsic spins depends on the magnitude of applied field which probably magnifies the spin-density fluctuations in the subsurface orbitals. Fluctuational mechanism of ``unpaired'' electron existence is consistent with the fact that nanodiamond has no unpaired electrons in absence of external field. 

Similar behavior of bulk materials subjected to mechanical damage was observed in Ge, Si, hexagonal SiC, diamond, MgO, CaO, ZnO, SrO, CaS and ZnS~\cite{Walters1961}. EPR line observed in these materials shows invariance under surface layer condition. It was found that ``paramagnetic centers do not involve impurities present in the starting material'', and authors concluded that ``it must be realized that powders prepared from diamagnetic starting materials might turn out to be strongly paramagnetic''~\cite{Walters1961}. Their observations are very similar to the nanodiamond case. Consideration of these results allows us to propose similar explanation for paramagnetic center appearance under mechanical damage as a result of spin-density fluctuations in long collective molecular orbital localized on the damaged area.

\section{Conclusion}

Three distinct classes of collective electron states have been found: collective bonding orbitals resembling the morphology of 3D-modulated particle in a box solutions; surface-localized non-bonding conductive Tamm states and subsurface-localized bonding states for non-uniformly compressed nanodiamond.

First-principles computations unambiguously show that every molecular orbital responsible for chemical bonding is collective. Surface compression plays crucial role in determining orbital localization inside nanodiamond, and we suggest to use ``Buried Layer Electron States'' term for these states.

On the base of buried layer electron states model we suggest that collective unpaired electrons are intrinsic to nanodiamond. Their subsurface localization is described in terms of surface compression arising form self-consistency of electrons and nuclei positions. Intrinsic spin existence is supposed to result from collective and spread nature of subsurface orbitals, allowing spin-density fluctuation effects to become significant on this length scale. Suggested model could explain free spins exhibited in experiments avoiding contradictions such as when one tries to attribute unpaired or free spin to radicals localized on nanodiamond surface.

Consideration of the density of states plot shows local increase in the subsurface-states density and shift of the valence and conduction band bottom towards lower energy. These results are promising and hopefully could explain peculiarity of the nanodiamond Auger spectra and PEELS experiments.

\section*{Abbreviations}
~\\
NMR~---~nuclear magnetic resonance; HOMO~---~highest occupied molecular orbital; LUMO~---~lowest unoccupied molecular orbital; EPR~---~electronic paramagnetic resonance; PEELS~---~parallel electron energy loss spectroscopy; PDB~---~protein database; RHF~---~restricted Hartree--Fock; DFT~---~density functional theory; LDA~---~local-density approximation; R-B3LYP~---~restricted Becke, three-parameter, Lee--Yang--Parr hybrid exchange-correlation functional.

\section*{Acknowledgments}

Thanks to the Genomic Research and Educational Center of SibFU for the provided access to the supercomputer cluster. The research was supported by the Ministry of Education and Science of the Russian Federation within the scope of mega-project ``Bioluminescent technology'' (contract No 11.G34.31.0058) by the rule No~220 at April 9th 2010 ``Measures to attract leading scientists in the Russian educational institution of higher education'' and government contract no. 14.A18.21.1911.


\begin{thebibliography}{10}
\expandafter\ifx\csname url\endcsname\relax
  \def\url#1{\texttt{#1}}\fi
\expandafter\ifx\csname urlprefix\endcsname\relax\def\urlprefix{URL }\fi
\providecommand{\bibinfo}[2]{#2}
\providecommand{\eprint}[2][]{\url{#2}}

\bibitem{Abrikosov1972}
\bibinfo{author}{Abrikosov, A.\,A.}
\newblock \emph{\bibinfo{title}{Introduction to the theory of normal metals}}
  (\bibinfo{publisher}{Academic Press}, \bibinfo{address}{New York and London},
  \bibinfo{year}{1972}).

\bibitem{Pate1986}
\bibinfo{author}{Pate, B.\,B.}
\newblock \bibinfo{title}{The diamond surface: atomic and electronic
  structure}.
\newblock \emph{\bibinfo{journal}{Surface science}}
  \textbf{\bibinfo{volume}{165}}, \bibinfo{pages}{83--142}
  (\bibinfo{year}{1986}).

\bibitem{Yang2002}
\bibinfo{author}{Yang, W.} \emph{et~al.}
\newblock \bibinfo{title}{{DNA-modified} nanocrystalline diamond thin-films as
  stable, biologically active substrates}.
\newblock \emph{\bibinfo{journal}{Nature Materials}}
  \textbf{\bibinfo{volume}{1}}, \bibinfo{pages}{253--257}
  (\bibinfo{year}{2002}).

\bibitem{Butler2008}
\bibinfo{author}{Butler, J.\,E.} \& \bibinfo{author}{Sumant, A.~V.}
\newblock \bibinfo{title}{The {CVD} of nanodiamond materials}.
\newblock \emph{\bibinfo{journal}{Chemical Vapor Deposition}}
  \textbf{\bibinfo{volume}{14}}, \bibinfo{pages}{145--160}
  (\bibinfo{year}{2008}).

\bibitem{Butler2009}
\bibinfo{author}{Butler, J.\,E.}, \bibinfo{author}{Mankelevich, Y.\,A.},
  \bibinfo{author}{Cheesman, A.}, \bibinfo{author}{Ma, J.} \&
  \bibinfo{author}{Ashfold, M.\,N.\,R.}
\newblock \bibinfo{title}{Understanding the chemical vapor deposition of
  diamond: recent progress}.
\newblock \emph{\bibinfo{journal}{Journal of Physics: Condensed Matter}}
  \textbf{\bibinfo{volume}{21}}, \bibinfo{pages}{364201}
  (\bibinfo{year}{2009}).

\bibitem{Balandin2011}
\bibinfo{author}{Balandin, A.\,A.}
\newblock \bibinfo{title}{Thermal properties of graphene and nanostructured
  carbon materials}.
\newblock \emph{\bibinfo{journal}{Nature Materials}}
  \textbf{\bibinfo{volume}{10}}, \bibinfo{pages}{569--581}
  (\bibinfo{year}{2011}).

\bibitem{Tamm1932g}
\bibinfo{author}{Tamm, I.\,E.}
\newblock \bibinfo{title}{\"Uber eine m\"ogliche art der elektronenbindung an
  kristalloberfl\"achen}.
\newblock \emph{\bibinfo{journal}{{Z.Phys.} Sowjetunion}}
  \textbf{\bibinfo{volume}{1}}, \bibinfo{pages}{733--746}
  (\bibinfo{year}{1932}).

\bibitem{Koutecky1957}
\bibinfo{author}{Kouteck\'y, J.}
\newblock \bibinfo{title}{Contribution to the theory of the surface electronic
  states in the {One-Electron} approximation}.
\newblock \emph{\bibinfo{journal}{Physical Review}}
  \textbf{\bibinfo{volume}{108}}, \bibinfo{pages}{13--18}
  (\bibinfo{year}{1957}).

\bibitem{Phariseau1960}
\bibinfo{author}{Phariseau, P.}
\newblock \bibinfo{title}{The energy spectrum of an amorphous substance}.
\newblock \emph{\bibinfo{journal}{Physica}} \textbf{\bibinfo{volume}{26}},
  \bibinfo{pages}{1185--1191} (\bibinfo{year}{1960}).

\bibitem{Bursill2001}
\bibinfo{author}{Bursill, L.\,A.}, \bibinfo{author}{Fullerton, A.\,L.} \&
  \bibinfo{author}{Bourgeois, L.~N.}
\newblock \bibinfo{title}{Size and surface structure of diamond nano-crystals}.
\newblock \emph{\bibinfo{journal}{International Journal of Modern Physics B}}
  \textbf{\bibinfo{volume}{15}}, \bibinfo{pages}{4087--4102}
  (\bibinfo{year}{2001}).

\bibitem{Peng2001}
\bibinfo{author}{Peng, J.}, \bibinfo{author}{Bulcock, S.},
  \bibinfo{author}{Belobrov, P.} \& \bibinfo{author}{Bursill, L.}
\newblock \bibinfo{title}{Surface bonding states of nano-crystalline diamond
  balls}.
\newblock \emph{\bibinfo{journal}{International Journal of Modern Physics B}}
  \textbf{\bibinfo{volume}{15}}, \bibinfo{pages}{4071--4086}
  (\bibinfo{year}{2001}).

\bibitem{Fang2009}
\bibinfo{author}{Fang, X.}, \bibinfo{author}{Mao, J.}, \bibinfo{author}{Levin,
  E.~M.} \& \bibinfo{author}{{Schmidt-Rohr}, K.}
\newblock \bibinfo{title}{Nonaromatic {Core-Shell} structure of nanodiamond
  from {Solid-State} {NMR} spectroscopy}.
\newblock \emph{\bibinfo{journal}{Journal of the American Chemical Society}}
  \textbf{\bibinfo{volume}{131}}, \bibinfo{pages}{1426--1435}
  (\bibinfo{year}{2009}).

\bibitem{Belobrov2003}
\bibinfo{author}{Belobrov, P.\,I.}, \bibinfo{author}{Bursill, L.\,A.},
  \bibinfo{author}{Maslakov, K.\,I.} \& \bibinfo{author}{Dementjev, A.\,P.}
\newblock \bibinfo{title}{Electron spectroscopy of nanodiamond surface states}.
\newblock \emph{\bibinfo{journal}{Applied Surface Science}}
  \textbf{\bibinfo{volume}{215}}, \bibinfo{pages}{169--177}
  (\bibinfo{year}{2003}).

\bibitem{Patrick2013}
\bibinfo{author}{Patrick, C.\,E.} \& \bibinfo{author}{Giustino, F.}
\newblock \bibinfo{title}{Quantum nuclear dynamics in the photophysics of
  diamondoids}.
\newblock \emph{\bibinfo{journal}{Nature Communications}}
  \textbf{\bibinfo{volume}{4}} (\bibinfo{year}{2013}).

\bibitem{Pfister2001}
\bibinfo{author}{Pfister, C.} \& \bibinfo{author}{Wirth, N.}
\newblock \bibinfo{title}{Component pascal language report}.
\newblock \bibinfo{type}{Tech. Rep.} \bibinfo{number}{2147483648},
  \bibinfo{address}{Z\"urich} (\bibinfo{year}{2001}).

\bibitem{Lide2009}
\bibinfo{author}{Lide, D.~R.} \emph{et~al.}
\newblock \emph{\bibinfo{title}{Handbook of Chemistry and Physics}}, vol.
  \bibinfo{volume}{131} (\bibinfo{publisher}{National Institute of Standards
  and Technology}, \bibinfo{year}{2009}), \bibinfo{edition}{90} edn.

\bibitem{Baerends1997}
\bibinfo{author}{Baerends, E.} \& \bibinfo{author}{Gritsenko, O.}
\newblock \bibinfo{title}{A quantum chemical view of density functional
  theory}.
\newblock \emph{\bibinfo{journal}{The Journal of Physical Chemistry A}}
  \textbf{\bibinfo{volume}{101}} (\bibinfo{year}{1997}).

\bibitem{Stowasser1999}
\bibinfo{author}{Stowasser, R.} \& \bibinfo{author}{Hoffmann, R.}
\newblock \bibinfo{title}{What do the {Kohn--Sham} orbitals and eigenvalues
  mean?}
\newblock \emph{\bibinfo{journal}{Journal of the American Chemical Society}}
  \textbf{\bibinfo{volume}{121}}, \bibinfo{pages}{3414--3420}
  (\bibinfo{year}{1999}).

\bibitem{Schmidt1993}
\bibinfo{author}{Schmidt, M.\,W.} \emph{et~al.}
\newblock \bibinfo{title}{General atomic and molecular electronic structure
  system}.
\newblock \emph{\bibinfo{journal}{Journal of Computational Chemistry}}
  \textbf{\bibinfo{volume}{14}}, \bibinfo{pages}{1347--1363}
  (\bibinfo{year}{1993}).

\bibitem{OBoyle2011}
\bibinfo{author}{{O'Boyle}, N.\,M.} \emph{et~al.}
\newblock \bibinfo{title}{Open babel: An open chemical toolbox.}
\newblock \emph{\bibinfo{journal}{Journal of cheminformatics}}
  \textbf{\bibinfo{volume}{3}}, \bibinfo{pages}{33--33} (\bibinfo{year}{2011}).

\bibitem{Hanwell2012}
\bibinfo{author}{Hanwell, M.\,D.} \emph{et~al.}
\newblock \bibinfo{title}{Avogadro: an advanced semantic chemical editor,
  visualization, and analysis platform.}
\newblock \emph{\bibinfo{journal}{Journal of cheminformatics}}
  \textbf{\bibinfo{volume}{4}}, \bibinfo{pages}{17--17} (\bibinfo{year}{2012}).

\bibitem{Halgren1996}
\bibinfo{author}{Halgren, T.} \& \bibinfo{author}{Nachbar, R.}
\newblock \bibinfo{title}{Merck molecular force field. {IV.} conformational
  energies and geometries for {MMFF94}}.
\newblock \emph{\bibinfo{journal}{Journal of Computational Chemistry}}
  \textbf{\bibinfo{volume}{17}}, \bibinfo{pages}{587--615}
  (\bibinfo{year}{1996}).

\bibitem{Humphrey1996}
\bibinfo{author}{Humphrey, W.}, \bibinfo{author}{Dalke, A.} \&
  \bibinfo{author}{Schulten, K.}
\newblock \bibinfo{title}{{VMD:} visual molecular dynamics}.
\newblock \emph{\bibinfo{journal}{Journal of Molecular Graphics}}
  \textbf{\bibinfo{volume}{14}}, \bibinfo{pages}{33--38}
  (\bibinfo{year}{1996}).

\bibitem{Ray2011}
\bibinfo{author}{Ray, M.} \emph{et~al.}
\newblock \bibinfo{title}{Towards electrons floating over diamond}.
\newblock \emph{\bibinfo{journal}{Bulletin of the American Physical Society}}
\newblock vol.~\bibinfo{volume}{56} (\bibinfo{year}{2011}).

\bibitem{Maier2000}
\bibinfo{author}{Maier, F.}, \bibinfo{author}{Riedel, M.},
  \bibinfo{author}{Mantel, B.}, \bibinfo{author}{Ristein, J.} \&
  \bibinfo{author}{Ley, L.}
\newblock \bibinfo{title}{Origin of surface conductivity in diamond}.
\newblock \emph{\bibinfo{journal}{Physical review letters}}
  \textbf{\bibinfo{volume}{85}}, \bibinfo{pages}{3472--5}
  (\bibinfo{year}{2000}).

\bibitem{Belobrov2001}
\bibinfo{author}{Belobrov, P.}, \bibinfo{author}{Gordeev, S.},
  \bibinfo{author}{Petrakovskaya, E.} \& \bibinfo{author}{Falaleev, O.}
\newblock \bibinfo{title}{Paramagnetic properties of nanodiamond}.
\newblock \emph{\bibinfo{journal}{Doklady Physics}}
  \textbf{\bibinfo{volume}{46}}, \bibinfo{pages}{459--462}
  (\bibinfo{year}{2001}).

\bibitem{Levin2008}
\bibinfo{author}{Levin, E.} \emph{et~al.}
\newblock \bibinfo{title}{Magnetization and c13 {NMR} spin-lattice relaxation
  of nanodiamond powder}.
\newblock \emph{\bibinfo{journal}{Physical Review B}}
  \textbf{\bibinfo{volume}{77}}, \bibinfo{pages}{1--10} (\bibinfo{year}{2008}).

\bibitem{Gordeev2013}
\bibinfo{author}{Gordeev, S.\,K.} \emph{et~al.}
\newblock \bibinfo{title}{Specific features in the change of electrical
  resistivity of carbon nanocomposites based on nanodiamonds under neutron
  irradiation}.
\newblock \emph{\bibinfo{journal}{Physics of the Solid State}}
  \textbf{\bibinfo{volume}{55}}, \bibinfo{pages}{1480--1486}
  (\bibinfo{year}{2013}).

\bibitem{Jiang2010}
\bibinfo{author}{Jiang, J.} \emph{et~al.}
\newblock \bibinfo{title}{Structure dependent quantum confinement effect in
  hydrogen-terminated nanodiamond clusters}.
\newblock \emph{\bibinfo{journal}{Journal of Applied Physics}}
  \textbf{\bibinfo{volume}{108}}, \bibinfo{pages}{094303--094303}
  (\bibinfo{year}{2010}).

\bibitem{Tian2009}
\bibinfo{author}{Tian, L.} \emph{et~al.}
\newblock \bibinfo{title}{{Carbon-Centered} free radicals in particulate matter
  emissions from wood and coal combustion.}
\newblock \emph{\bibinfo{journal}{Energy \& fuels : an American Chemical
  Society journal}} \textbf{\bibinfo{volume}{23}}, \bibinfo{pages}{2523--2526}
  (\bibinfo{year}{2009}).

\bibitem{Panich2012}
\bibinfo{author}{Panich, A.\,M.}
\newblock \bibinfo{title}{Nuclear magnetic resonance studies of nanodiamonds}.
\newblock \emph{\bibinfo{journal}{Critical Reviews in Solid State and Materials
  Sciences}} \textbf{\bibinfo{volume}{37}}, \bibinfo{pages}{276--303}
  (\bibinfo{year}{2012}).

\bibitem{Fukutome1968}
\bibinfo{author}{Fukutome, H}
\newblock \bibinfo{title}{Spin Density Wave and Charge Transfer Wave in Long Conjugated Molecules Molecules}.
\newblock \emph{\bibinfo{journal}{Progress of Theoretical Physics}} \textbf{\bibinfo{volume}{40}}, \bibinfo{pages}{1227--1245}
  (\bibinfo{year}{1968}).


\bibitem{Walters1961}
\bibinfo{author}{Walters, G.\,K.} \emph{et~al.}
\newblock \bibinfo{title}{Paramagnetic Resonance of Defects Introduced Near the Surface of Solids by Mechanical Damage}.
\newblock \emph{\bibinfo{journal}{Journal of Applied Physics}} \textbf{\bibinfo{volume}{32}}, \bibinfo{pages}{1854--1859}
  (\bibinfo{year}{1961}).
  
\end{thebibliography}

\end{document}